\documentclass{article}
\usepackage{authblk}
\usepackage{commath}
\usepackage{mathtools}
\usepackage{amsmath}
\usepackage{amssymb}
\usepackage[margin=0.8in]{geometry}
\usepackage{braket}
\usepackage{physics}
\usepackage{graphicx}
\usepackage{blindtext}
\usepackage[utf8]{inputenc}
\usepackage[many]{tcolorbox}
\usepackage[useregional]{datetime2}

\newcommand*\diff{\mathop{}\!\mathrm{d}}

\newcommand{\mydate}{\DTMdisplaydate{2023}{11}{28}{-1}}

\title{Deriving the one-electron Spectral Function for the 1D Hubbard Model}
\date{\vspace{-5ex}}
\author{Daniel Bozi}
\affil{Institute for pedagogical sciences, Högskolan i Borås, 501 90 Sweden \protect \\ Email: daniel.bozi@hb.se}

\begin{document}
\maketitle
\quad \quad \quad \quad \quad \quad \quad \quad \quad \quad \quad \quad \quad \quad \quad \quad \quad \quad \quad \quad \quad \mydate

\begin{abstract}
This pre-print deals with the one dimensional Hubbard model, as described by the Pseudofermion Dynamical Theory (PDT), with the purpose of (1) deriving a novel expression for the one electron spectral function for all values of the on-site repulsion $U/t$ and filling $n \in (0,1)$, at vanishing magnetisation $m \rightarrow 0$, and (2) discover how to correctly compare the results originating from two different theoretical frameworks in the $U/t \rightarrow \infty$ limit, as a first-test of the novel expressions obtained in this paper. Both purposes are fulfilled; an exact expression of the spectral function is obtained, which is successfully compared with previously known results in the $U \rightarrow \infty$ limit. 
\newline

Following the PDT, the expression for the one electron spectral function factorises into a spin part and a charge part for all values of the on-site repulsion $U$, where the quantum objects are spin zero and $\eta$-spin (charge) zero singlet pairs of so-called rotated electrons, which in turn are obtainable from the original electrons by a unitary transformation. The spectral function thus obtained is exemplified for $U/t = 400$, with the purpose of comparing it with the same function obtained by other authors (and other means) in the $U \rightarrow \infty$ limit. 
\newline

The main pillars of the PDT is presented in a summarised form. For example, we will only be interested in excited energy eigenstates which originate the overall largest weight to the spectral map in the $(k,\omega)$ plane, safely ignoring higher order contributions. Even though emphasis is given on step-by-step derivations where necessary, derivations that have been done elsewhere and/or do not notably contribute to the physical understanding, are sometimes avoided. Therefore, references for further study are given throughout the paper.

\end{abstract}
\newpage

\tableofcontents

\section{Introduction}

\subsection{Why the Hubbard model?}
The one-dimensional Hubbard model, introduced by Hubbard in 1963 \cite{hubbard1}, stands as an enduring paradigm in condensed matter physics, providing a theoretical framework to describe and analyse the electron-electron interactions and quantum correlations in one-dimensional systems. Originally conceived to capture the essence of electron correlations in strongly correlated systems, the model has emerged as a remarkable tool for understanding the electronic behavior of quasi one-dimensional materials, with its utility lying in its capacity to capture the essence of electronic correlations, where electron-electron interactions play a pivotal role in determining a material's electronic and magnetic properties. Over the years, the model has evolved into a cornerstone of condensed matter theory, serving as a theoretical laboratory for investigating phenomena such as Mott insulator transitions \cite{mott0}-\cite{mott5}, spin-charge separation \cite{scsep1}-\cite{scsep5}, as well as superconducting phases \cite{supercond1}-\cite{supercond5}. A pedagogical starting point in understanding why standard Landau theory can not be applied in such one dimensional systems is the realisation that the low-lying energy excitations do not follow the usual Fermi liquid description, but are instead replaced by another theoretical framework, namely the Tomonaga-Luttinger Liquid (TLL) \cite{TLL1}. In this framework, the low-lying energy excitations form collective motions of bosons to the contrary of the usual quasiparticle description \cite{TLL2}-\cite{TLL4}. The one-dimensional Hubbard model serves as a fundamental theoretical tool for studying TLL behavior, providing a mathematical representation of how electrons with repulsive interactions propagate through a 1D lattice, revealing charge density fluctuations and spin-charge separation, as indeed characterised by the Tomonaga-Luttinger liquid \cite{scsep6}-\cite{scsep10}. But the justification for studying the one dimensional Hubbard model is not merely theoretical, as a plethora of experimental breakthroughs have underscored the Hubbard model's relevance in capturing the intricate electronic behavior of quasi one-dimensional materials. Most notably, experiments involving ultracold atomic gases confined to optical lattices have opened new avenues for exploring quantum many-body physics \cite{ultracold1}-\cite{ultracold5} while the realization of Mott insulator transitions and the observation of correlated states in ultracold atomic systems provide a compelling bridge between theory and experiment \cite{ultracold6}.

\subsection{The theoretical framework used in this paper}
The theoretical foundation of the results presented in this paper is given by (1) the work done by Lieb and Wu in their seminal paper from 1968 \cite{LiebWu}, where by using a Bethe Ansatz approach, the problem of diagonalising the 1D Hubbard model hamiltonian was reduced to solving a set of coupled non-linear equations, and (2) the Takahashi string hypothesis from 1972, in which the quantum numbers introduced by the Lieb-Wu equations were organised into so-called strings \cite{takahashi}, connected to each other via the \textit{thermodynamic Takahashi equations} (these equations are quite lengthy and not directly used in this paper, hence they are omitted from this brief overview). The quantum numbers are dubbed the (real) \textit{charge momenta} and the (complex) \textit{spin rapidities} respectively, illustrating the spin-charge separation for which the Hubbard model is so famous. In a long succession of publications starting from the late 1990's, J. M. P. Carmelo and others have connected these quantum objects to the original electrons of the hamiltonian, constituting the mathematical basis for the pseudofermion picture used here below. For very large system sizes (as measured by the length of the 1D lattice $L=a N_a$ where $a$ is the lattice constant and $N_a$ the number of lattice sites), the pseudofermion description is valid for the entire energy excitation spectrum \cite{josenuno}. They obey Haldane statistics \cite{haldane} \cite{josenuno}, and are spin and $\eta$-spin zero objects, corresponding to spin- and $\eta$-spin singlet pairs of the $+1/2$ and $-1/2$ spin and $\eta$-spin projections, respectively, of so-called \textit{rotated electrons} also introduced here below. As an energy eigenstate can be described by the occupancy configurations of the Takahashi quantum numbers, energy eigenstates are also completely described by occupancy configurations of pseudofermions, given that the latter stem from the former, as the pseudofermions are the dynamical scatterers and scattering centers of the model. The \text{pseudofermion dynamical theory} (PDT) has been presented in numerous publications by J. M. P. Carmelo and others \cite{carmelo1}-\cite{carmelo61} (note particularly the comprehensive review of Ref. \cite{carmelo62}), and is often explained in the context of experimental results relating to one- and two-electron correlation functions \cite{carmelo7}-\cite{carmelo10}. 
\newline

\section{The model hamiltonian and the rotated electron}
Let us consider a 1D Hubbard model on a periodic lattice

\begin{equation}
\hat{H} = -t \sum_{\langle i,j \rangle \sigma} c^{\dagger}_{i, \sigma} c_{j,\sigma} + U \sum_i n_{i,\uparrow} n_{i,\downarrow},
\end{equation}

where the summation $\langle i,j \rangle$ is done over nearest neighbour lattice sites and where it is understood that the $\langle i,i \rangle$ term is absorbed into the chemical potential in a grand canonical ensemble description. $t$ is the transfer integral (also known as the nearest-neighbour hopping strength), and $U$ is the effective on-site Coulomb repulsion, a number which will always be larger than zero. Furthermore, we let the electron filling $n=N/L \in (0,1)$ where $N$ is the number of electrons and where such units are chosen so that the lattice constant $a=1$ (the corresponding system with $n\in (1,2)$ follows an equivalent description by performing an $\eta$ spin-flip on the hamiltonian). At half-filling, $n=1$, the Hubbard hamiltonian is mott-insulating \cite{halff1} for all values of $U/t>0$ while the framework presented here focuses on the dynamical metallic phase \cite{carmelo7} \cite{carmelo9}, rendering $0<n<1$. Similarily, we let the overall magnetisation of the system be $m>0$ but vanishing, $m=( N_{\uparrow} - N_{\downarrow} ) / L \rightarrow 0$, where $N_{\sigma}$ are the total number of electrons with spin projection $\sigma$. 
\newline

The bridge between the original electron description of the model hamiltonian and the pseudofermions mentioned above, is constituted by a unitary transformation that makes double occupancy $D=\langle \sum_i c^{\dagger}_{i,\uparrow} c_{i,\uparrow} c^{\dagger}_{i,\downarrow} c_{i,\downarrow} \rangle$ a good quantum number for all values of the on-site repulsion $U$ (for electrons, $D$ is only a good quantum number for $U=\infty$) \cite{carmelo3}, \cite{carmelo61} - \cite{carmelo64}. This unitary transformation, for $U/t \rightarrow \infty$, cancels all terms in the original hamiltonian that change the number of doubly occupied sites, enabling a reformulation of the model hamiltonian $\hat{H} = \hat{H}^{(0)} + \hat{H}^{(1)} + \hat{H}^{(2)} + \ldots$, where $\hat{H}^{(j)}$ has precisely $j$ number of doubly occupied sites \cite{rotel1}, \cite{rotel2}. This means that, in our present framework, we map our finite-$U$ model onto a transformed one which "resembles" the $U=\infty$ model, however with some key differences. For example, in this limit, any eigenstate of the model is usually factorized into two eigenstates: one describing spinless fermions and another describing chargeless spins \cite{Uinfty1}, where the spin part maps onto a 1D antiferromagnetic Heisenberg spin hamiltonian \cite{Uinfty2}. In the framework described here however, even though the wave function still factorises in a similar way, quantum objects remain the ones originally derived from the string hypothesis, most notably with spin and $\eta$-spin projection zero as we shall see below. However, due to the $U=\infty$ history of the unitary transformation, comparison of our results with known $U=\infty$ results will be a crucial test of the framework used here. Example of such results include use of the wave-function factorization to calculate one-electron correlation functions \cite{penc1}-\cite{penc4}.
\newline

Now, all meaningful physics of the theoretical framework described in the previous section hinges on the unitary transformation $\hat{V}$ which "rotates" the electron creation and annihilation operators, rendering the energy eigenstates of the hamiltonian to be described by occupancy configurations of so-called \textit{rotated electrons}. The corresponding "rotated hamiltonian" $\tilde{H} = \hat{V} \hat{H} \hat{V}^{\dagger}$ can then be expressed in terms of the rotated electron creation and annihilation operators $\tilde{c}_{i\sigma}^{\dagger}$ and $\tilde{c}_{i\sigma}$, respectively, where a rotated electron creatoin and annihilation operator acting on lattice site $i$ is given by: 

\begin{equation}
\tilde{c}_{i\sigma}^{\dagger} = \hat{V} c_{i\sigma}^{\dagger} \hat{V}^{\dagger} \ , \qquad \tilde{c}_{i\sigma} = \hat{V} c_{i\sigma} \hat{V}^{\dagger}
\end{equation}

where $\hat{V} \hat{V}^{\dagger} = \mathbf{1}$ on account of being unitary. Here it should be noted that the total momentum operator $\hat{P}$ commutes with the unitary transformation operator, rendering the original electron lattice and the rotated lattice identical. That this is the case is simply by construction: since double occupancy of rotated electrons is a good quantum number for all values of $U/t$, then so are the single spin-$\downarrow$ and single spin-$\uparrow$ occupancies, as well as no-occupancies (empty sites), due to the fact that these numbers are not linearly independent (they are, in order, $N_{\downarrow}-D$, $N-N_{\downarrow}-D$, and $N_a-N+D$, respectively). This unitary transformation can be described in terms of expansion in powers of $t/U$, where indeed the hamiltonian $\hat{H}^{(j)}$ is of the order of $(t/U)^j$ (the $t/U$ expansion of the Hubbard model was explicitly calculated up to eigth order in Ref. \cite{rotel2} using this transformation). This expansion is made possible by the Baker-Hausdorff Lemma by realising that any unitary operator $\hat{V}$ can be written as the exponential of an anti-hermitian operator. However, we do not need the series expansion to prove the existence, uniqueness and unitariness of $\hat{V}$ \cite{carmelo3} \cite{carmelo62}. 

\section{Pseudofermions}

\subsection{... but first, $c$ and $s$ particles}

The thermodynamic behaviour of the 1D Hubbard hamiltonian has been described by use of the Takahashi quantum numbers for more than 30 years \cite{others1} - \cite{others3}. Usually, however, the quantum objects emerging from this description have been ”spinons” and ”holons” with $\eta$-spin and spin projection equal to $\pm 1/2$, whilst here, the scatterers and the scattering centers are both spinless and $\eta$-spinless \cite{carmeloS}. The PDT referenced above claims to provide the missing link between the original electrons and the quantum objects diagonalising the Hubbard hamiltonian, by the unitary transformation which maps any $U/t>0$ Hubbard hamiltonian onto a "$U=\infty$ similar hamiltonian", where occupancy numbers of double, single, and no-occupancy are all good quantum numbers for the occupancy configurations of the rotated electrons. Furthermore, the charge and spin degrees of freedom of these rotated electrons separate from each other, and recombine in new quantum objects called the $c$ and $s$ particles. 
\newline

Here we summarise these particle's microscopic constituents, further explained in \cite{carmelo61}-\cite{carmelo64}: 
\newline

\begin{enumerate}
  \item \textit{c-particle}: an $\eta$-spin singlet pair of the charge degree of freedom of a singly occupied rotated electron and a singly occupied rotated electron hole, 
  \item \textit{s-particle}: a spin singlet $(\uparrow,\downarrow)$ pair of the spin degree of freedom of two singly occupied rotated electrons, and 
  \item $\alpha n$ particles: $n=1, 2, \ldots$ number of pairs of the charge degrees of freedom on doubly occupied and empty sites ($\alpha=c$), or $n= 2, 3, \ldots$ number of spin singlet pairs of the spin degree of freedom on singly occupied sites ($\alpha=s$). 
\end{enumerate}

It should be noted that in what follows of this paper, we will omit particles of type (3) since the ground state of the model only contain particles of type (1) and (2), and their appearence in excited energy eigenstates only have marginal impact on the sum rules for the one-electron addition and removal spectral function \cite{josenuno} \cite{carmelo94percent}.
Considering that we assume $m>0$, it follows that $N_{\uparrow} > N_{\downarrow}$ meaning that some rotated electron spins with spin projection $+1/2$ will remain unpaired when forming the $s$ particles. These unpaired quantum objects do not participate in the dynamics of the model \cite{carmelo3} but all remaining states of the Hilbert space are attainable by acting on these unpaired quantum objects by spin raising operators as expressed in terms of creation and annihilation operators of rotated electrons: $\hat{S}_s^+ = \sum_i \tilde{c}_{i,\downarrow}^{\dagger} \tilde{c}_{i,\uparrow}$. For example, where all singly occupied $\downarrow$-spin rotated electrons have been paired with an equal number of $\uparrow$-spins in order to form $s$-particles, the remaining unpaired rotated electron $\uparrow$-spins contribute to the total spin projection $S_s^z = \langle \hat{N_{\uparrow}} - \hat{N_{\downarrow}} \rangle /2$ of the system. This kind of eigenstate is a \textit{lowest weight state}, and the remaining $2S_s$ eigenstates describable by the model are then reached by successive application of the spin raising operator above. This reasoning is equivalent for the $\eta$-spin channel, in turn forming a tower consisting of $2S_c + 1$ states. It has been shown that the Bethe ansatz solution only accounts for either lowest (or highest) weight states of the one dimensional Hubbard model \cite{others1} \cite{others11}. However, after taking into account all the states reached by the off-diagonal generators of the spin- and the $\eta$-spin algebras, the Bethe ansatz solution is indeed complete, in that the total number of states present in the solution gives the accurate dimension of the Hilbert space of the original model \cite{others12}. In the following, we shall only consider such lowest weight states, where the number of $\downarrow$-spins equate the numbers of $s$ particles, and the number of rotated electrons $N$ equate the number of $c$ particles (hence no doubly occupied sites). 
\newline

\subsection{Phase shift of $c$ and $s$ particle momentum}

This paper will be primarily focused on the one-electron spectral function. It will hence be natural for us to discuss excited energy eigenstates in the context of one-electron removal, or one-electron addition, respectively. As we shall see, this is evidenced in the Lehman representation of the one-particle spectral functions, evaluating overlaps between an $N$ particle ground state, and an $N\pm1$ particle excited final state. Eigenstates in such excited systems are characterised by a phase they pick up at each scattering event, where the pseudofermions are both the active scatterers and scattering centers of the dynamical theory. It is this non-zero phase shift which is responsible for all dynamical properties of the correlation function, and the origin of said phase shift can once again be found in the continuous Takahashi equations. This phase shift is also what differs between the $c$ and $s$ particles described above on the one hand, and the pseudofermions described below on the other. As we shall justify below,  phase shifts describe scattering events of a zero-energy forward scattering type, which is a crucial property of these quantum objects, enabling a factorisation of the wave function for all values of $U/t$. In order to avoid confusion between the $c$ and $s$ particles discussed above and the following pseudofermions, we shall note that both of these quantum objects are described in terms of rotated electrons in the exact same way; the only change between these two is the two-particle phase shift, which for the $c$ and $s$ particles is zero.
\newline

The ground state is characterised by a densely packed configuration of the Takahashi quantum numbers. The largest and smallest of such quantum numbers define the Brillouin zones of the particle bands: each rotated electron lattice site can host a $c$ particle, hence the Brillouin zone for this particle is defined by the number of sites: $q_{c}^{\pm} =\pm (2\pi /L) \cdot (L-1)/2 = \pm (\pi-\pi/L)$, while each $\uparrow$-spin can pair with a $\downarrow$-spin, meaning that the Brillouin zone for the $s$ particles is given by the number of $\uparrow$-spins: $q_{s}^{\pm}=\pm (2\pi /L) \cdot (N_{\uparrow}-1)/2 = \pm (\pi n_{\uparrow}-\pi/L)$. The actual number of $c$ particles is nothing but the number of electrons $N$, which in turn defines the Fermi boundary for the $c$ particle momenta $q_{Fc}^{\pm}=\pm (2\pi /L) \cdot (N-1)/2 = \pm (\pi n -\pi/L)$, and lastly, the number of $\downarrow$-spins governs how many $s$ particles we can form, i.e. $q_{Fs}^{\pm}=\pm (2\pi /L) \cdot (N_{\downarrow}-1)/2 = \pm (\pi n_{\downarrow}-\pi/L)$. Notice that at zero magnetisation, $m=0$, we have that $n_{\uparrow} = n_{\downarrow}$ and the $s$-band is full in that case. 
\newline

This choice of number of rotated electron sites and number of rotated electrons is not unique. In fact, there are many choices for these numbers, rendering the Takahashi quantum numbers being either integers or half-odd integers. In our case, we have let, for an integer $l$, $N_a=4l+2$ be an even number of $c$ pseudofermion lattice sites, with the odd numbers $N_{\downarrow}$ and $N_{\uparrow}$ denote the number of rotated electron down- and up-spins, respectively. This in turn guarantees an even number $N$ of rotated electrons (and electrons), as $N=N_{\downarrow}+N_{\uparrow}$, with $N_{\uparrow}$ being the number of lattice sites for the $s$ pseudofermion. The complete set of choices for these numbers (and implications for the definitions of the Brillouin zones and Fermi momenta) are listed, for example, in Ref. \cite{carmelo3}.
\newline

The fact that we describe eigenstates in this way means that a state reachable by our ground state (in this paper relevant in the context of the one electron spectral function) might be described by quantum numbers different in kind from that of the ground state (going from being either integers to half-odd integers or vice versa). This means that the entire sea of particles in the excited energy eigenstate might have shifted momenta values corresponding to the shift in the quantum numbers. When this happens it will be valid for all quantum numbers, resulting in a current either to the right ($+\pi/L$) or to the left ($-\pi/L$) of all momenta values. In our application, this global shift occurs in the $c$ band if the change in the number $N_a/2 - N_{\downarrow}$ is odd due to the ground state $\rightarrow$ excited final state transition, while the same occurs in the $s$ band if the change in the number $N_{\uparrow}$ is odd.
\newline

In the ground state, the number occupancy configuration for these particles will be given by the Heaviside $\theta$-function: $N_{\alpha}^0 (q) = \theta(q_{F\alpha} - |q|)$, where $\alpha=c$ or $s$. When adding or removing an electron, a typical energy eigenstate will have a deviated occupancy configuration of these particles, rendering an occupancy configuration $N_{\alpha} (q) = N_{\alpha}^0 (q) + \Delta N_{\alpha} (q)$ (how addition and removal of an electron affects occupancy configurations of $c$ and $s$ pseudofermions will be discussed below). This leads to an energy deviation $\Delta E$ relative to the ground state energy $E_{GS}$, defining the $\alpha = c$ and $s$ particle energy bands with dispersion relation given by $\varepsilon_{\alpha}$, as well as higher order residual energy terms containing mixed terms of type $\Delta N_{\alpha} (q) \Delta N_{\alpha'} (q')$ which we will not write down explicity:

\begin{equation}
\begin{split}
\Delta E &= \sum_{\alpha = c, s} \sum_{q_{\alpha}} \varepsilon_{\alpha} (q_{\alpha}) \Delta N_{\alpha} (q_{\alpha}) + E (\text{res. int.})\\
\varepsilon_{\alpha} (q_{\alpha}) &= \frac {\delta \Delta E} {\delta \Delta N_{\alpha} (q_{\alpha})} \label{energybands}
\end{split}
\end{equation}

where the sum is over the entire Brillouin zone and $\delta$ signifies the functional derivate taken in the thermodynamic limit, where conversion from sums to integrals makes use of the $c$ and $s$ particle spacing $q_{i+1} - q_i = 2\pi /L$.
Having these energy bands defined over the entire energy excitation spectrum makes it possible to calculate the one-electron spectral function at all excitation energies. Now, using the thermodynamic Takahashi equations relevant in this context:
\begin{equation}
\begin{split}
k(q) &= q + \frac 1 {\pi} \int_{q_s^-}^{q_s^+} \diff q' N_s (q') \arctan (\frac {\Lambda_s (q') - \sin k(q)} u) \\ \\
q &= \frac 1 {\pi} \int_{q_c^-}^{q_c^+} \diff q' N_c (q') \arctan (\frac {\Lambda_s (q) - \sin k(q')} u) + \frac 1 {\pi} \int_{q_s^-}^{q_s^+} \diff q' N_s (q') \arctan (\frac {\Lambda_s (q') - \Lambda_s (q)} {2u})
\end{split} \label{eqtakahashi}
\end{equation}

where $k(q)$ and $\Lambda_s (q)$ are Takahashi's rapidity momentum $k_j$ and spin rapidity $\Lambda_j$ in the thermodynamic limit, hereby known as \textit{rapidity functions}, with the momentum $q_j=(2\pi/L) I_j^{\alpha}$ for the quantum numbers $I_j^{\alpha}$ of the $\alpha = c,s$ kinds introduced by Takahashi. 
\newline

In this limit, the total energy can be written as $E=E_{GS} + \Delta E + E(\text{res.int.})$, where $\Delta E$ is given by Eq. (\ref{energybands}), and where the ground state energy is given by
\begin{equation}
E_{GS} = -t \frac L {\pi} \int_{q_c^-}^{q_c^+} N_c (q) \cos k(q). \label{EGS}
\end{equation}

Introducing the occupancy deviations $N_{\alpha} (q) = N_{\alpha}^0 (q) + \Delta N_{\alpha} (q)$ into Eqs. (\ref{eqtakahashi}) and (\ref{EGS}) and equating order by order, one straightforwardly obtain expressions for (1) the corresponding deviations in the rapidity functions (i.e. by defining $k(q) = k^0 (q) + \Delta k(q)$, where $k^0(q)$ is the ground state rapidity function, and similarily for $\Lambda_s (q)$), and (2) the expressions for each pseudofermion energy dispersion relation, expressed as integral equations involving phase shifts as further specified below. In the following, we follow the results of Ref. \cite{josenuno2}, by noting that the equivalence between the following two expansions can be used to define a phase shift $\Phi_c (q)$ in the thermodynamic limit:

\begin{equation}
\begin{split}
                \left.
                \begin{alignedat}{4}
                    &k(q) = k^0 (q) + \Delta k(q) = k^0 (q) + \frac {dk^0 (q)} {dq} \Delta Q_c (q)  \quad \\
                    &k^0 (q+\delta(q)) = k^0 (q) + \frac {dk^0 (q)} {dq} \delta (q) + \ldots \\
                \end{alignedat}
                 \right\}
                &\qquad \implies \delta (q) = \Delta Q_c (q) = \frac {2\pi} L \Phi_c (q)
                \\
\end{split}
\end{equation}

\textit{defining} the function $\Phi_c (q)$ (equivalently for $\Phi_s (q)$). This means that a shift in the ground state momentum of a $c$ or $s$ particle, results in an excited energy eigenstate described by the ground state rapidity function, in other words, $k(q) = k^0 (\bar{q})$, where $\bar{q} = q + (2\pi / L) \Phi_c (q)$, meaning that pseudofermions can be viewed as having a non-interacting character (due to allowing excited final states being described by ground state rapidity functions). By introducing this expression into Eq. (\ref{eqtakahashi}) one can rewrite the rapidity deviations in terms of so-called dressed phase shifts $\Phi_{\alpha, \alpha'} (q,q')$ if we define these according to:

\begin{equation}
\begin{split}
\Phi_{\alpha} (q_{\alpha}) &= \sum_{\alpha' = c, s} \sum_{q_{\alpha'}} \Phi_{\alpha, \alpha'} (q_{\alpha},q_{\alpha'}) \Delta N_{\alpha'} (q_{\alpha'}) \\
\Phi_{\alpha \alpha'} (q_{\alpha},q_{\alpha'}) &= \frac {\delta \Phi_{\alpha} (q_{\alpha})} {\delta \Delta N_{\alpha'} (q_{\alpha'})}
\end{split}
\end{equation}

\begin{figure}[h!]
  \includegraphics[width=0.99\linewidth]{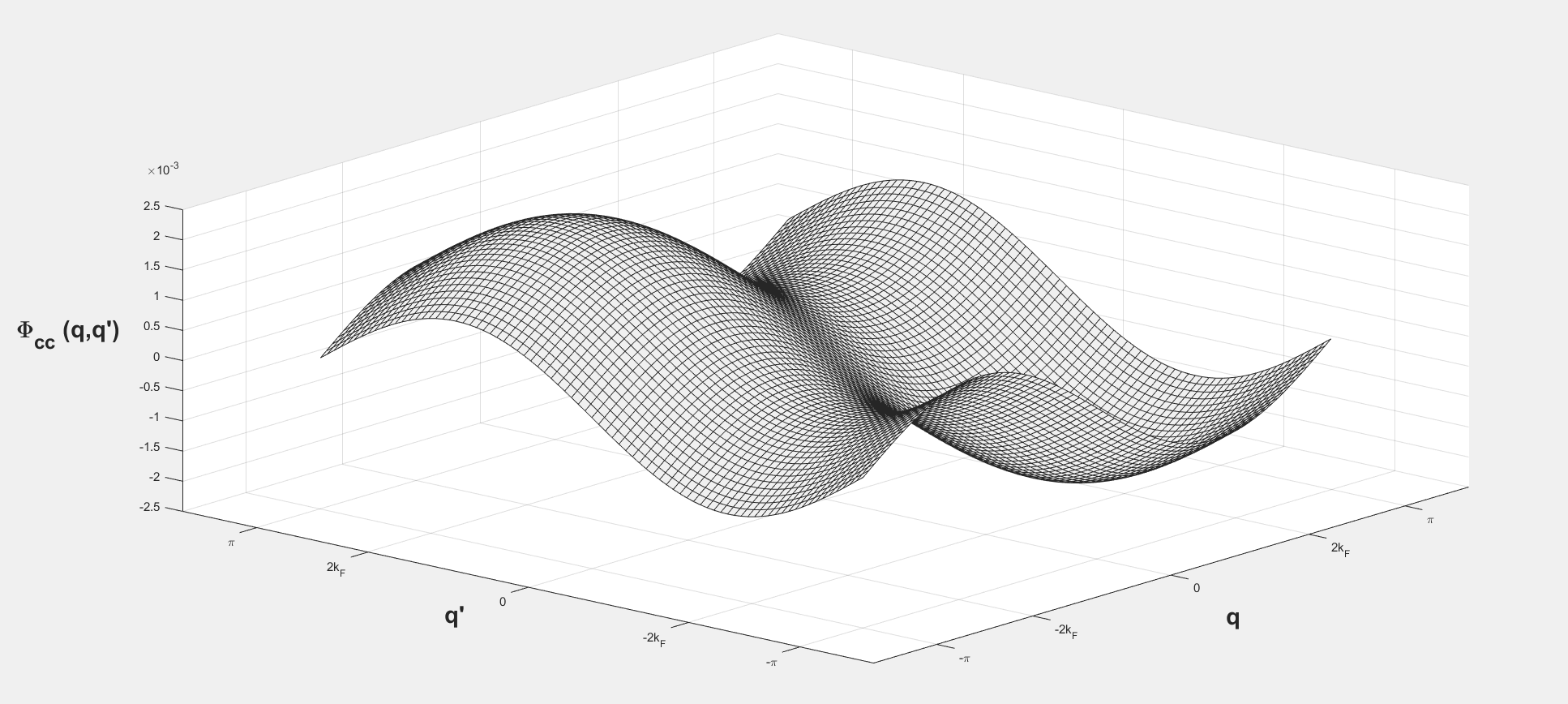}
  \centering
  \caption{The phase shift $\Phi_{cc} (q,q')$ in the $U/t \rightarrow \infty$ limit, for filling $n=0.59$. The filling $n=0.59$ is chosen from an experimental application regarding the charge transfer salt TTF-TCNQ studied in Ref. \cite{TTFTCNQ}.}
  \label{phicc}
\end{figure}

\begin{figure}[h!]
  \includegraphics[width=0.99\linewidth]{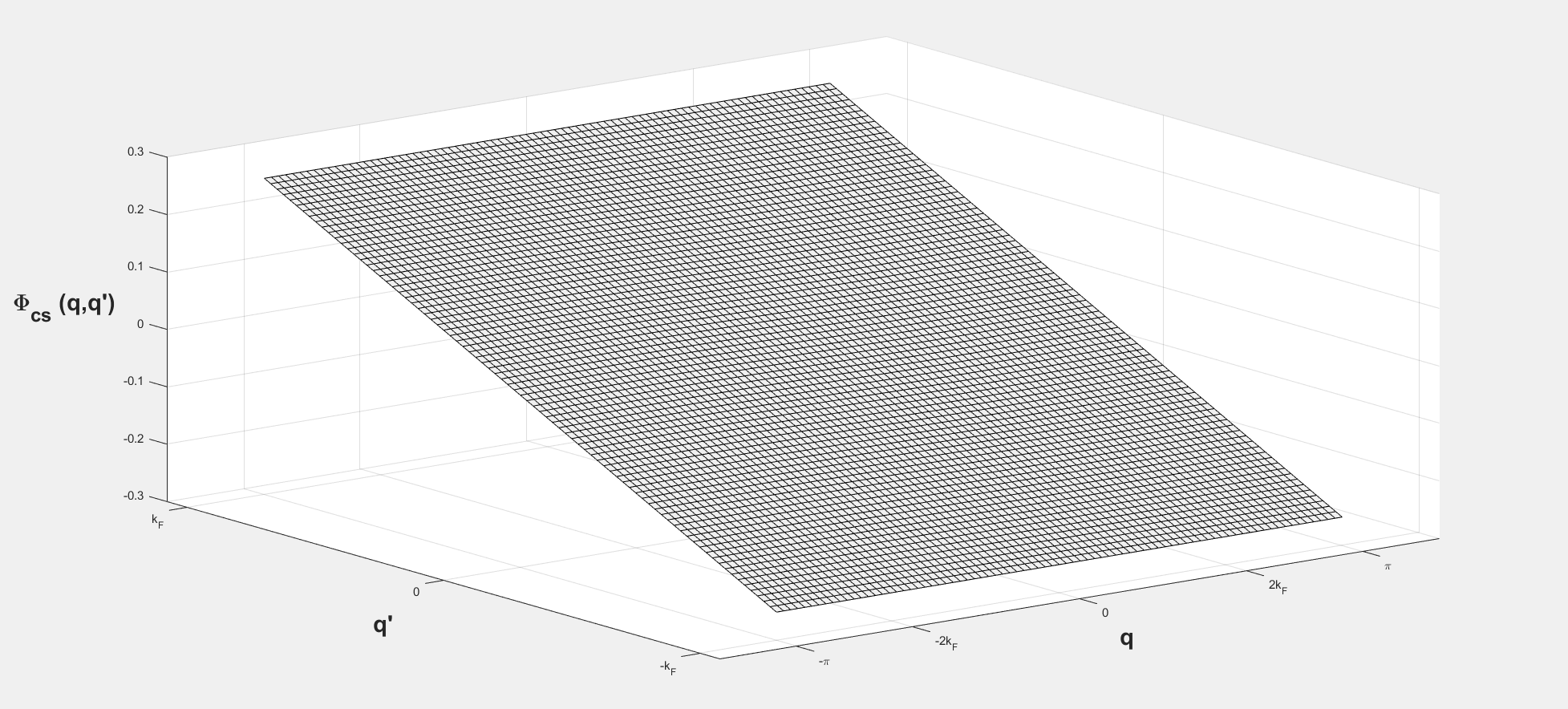}
  \centering
  \caption{The phase shift $\Phi_{cs} (q,q')$ in the $U/t \rightarrow \infty$ limit, for filling $n=0.59$. The filling $n=0.59$ is chosen from an experimental application regarding the charge transfer salt TTF-TCNQ studied in Ref. \cite{TTFTCNQ}.}
  \label{phics}
\end{figure}

\begin{figure}[h!]
  \includegraphics[width=0.99\linewidth]{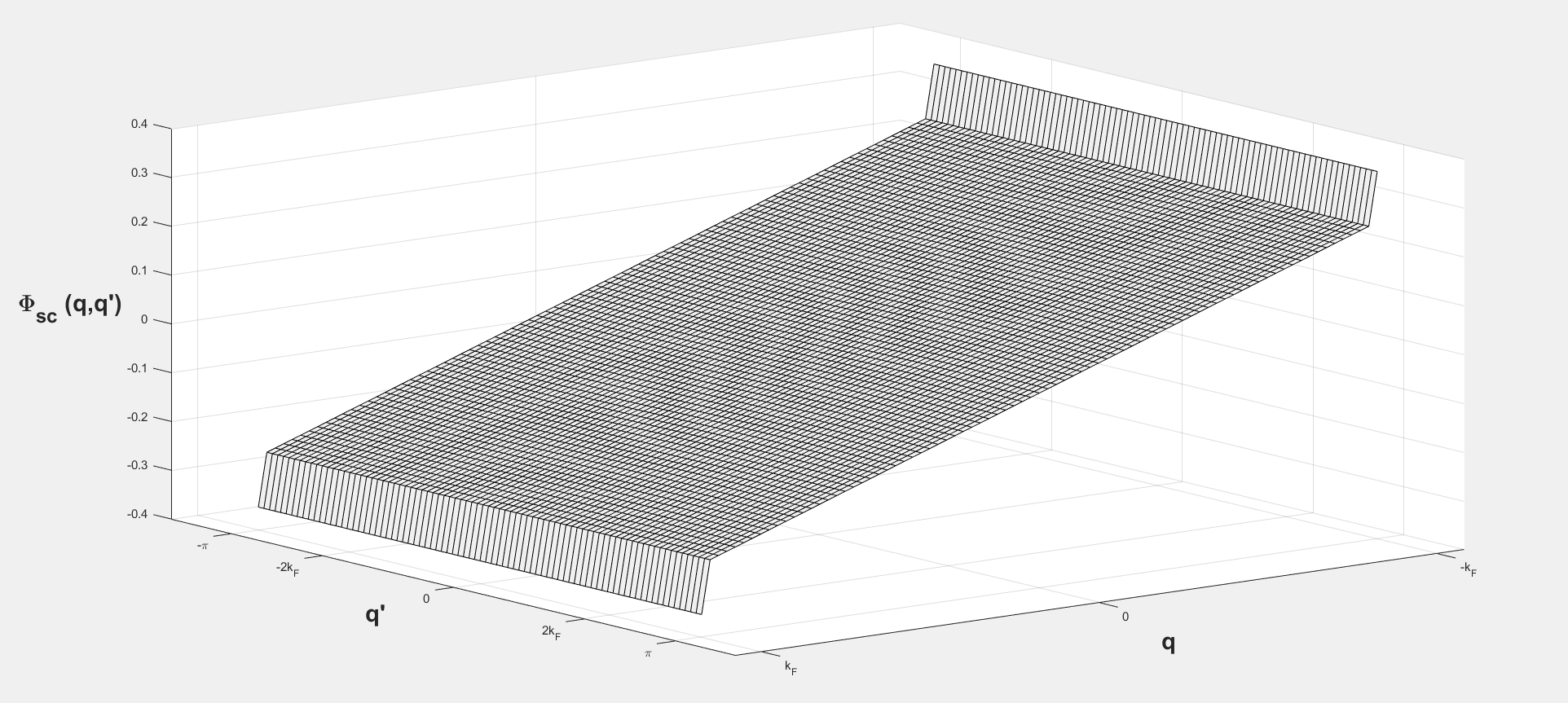}
  \centering
  \caption{The phase shift $\Phi_{sc} (q,q')$ in the $U/t \rightarrow \infty$ limit, for filling $n=0.59$. The filling $n=0.59$ is chosen from an experimental application regarding the charge transfer salt TTF-TCNQ studied in Ref. \cite{TTFTCNQ}. Note the discontinuity at the Fermi momentum $q=-k_F$ and $q=k_F$, which is a $m \rightarrow 0$ feature. In numerical applications below, we shall allow a small but vanishing magnetisation to avoid singular behaviors in the derivative of this phase shift.}
  \label{phisc}
\end{figure}

\begin{figure}[h!]
  \includegraphics[width=0.99\linewidth]{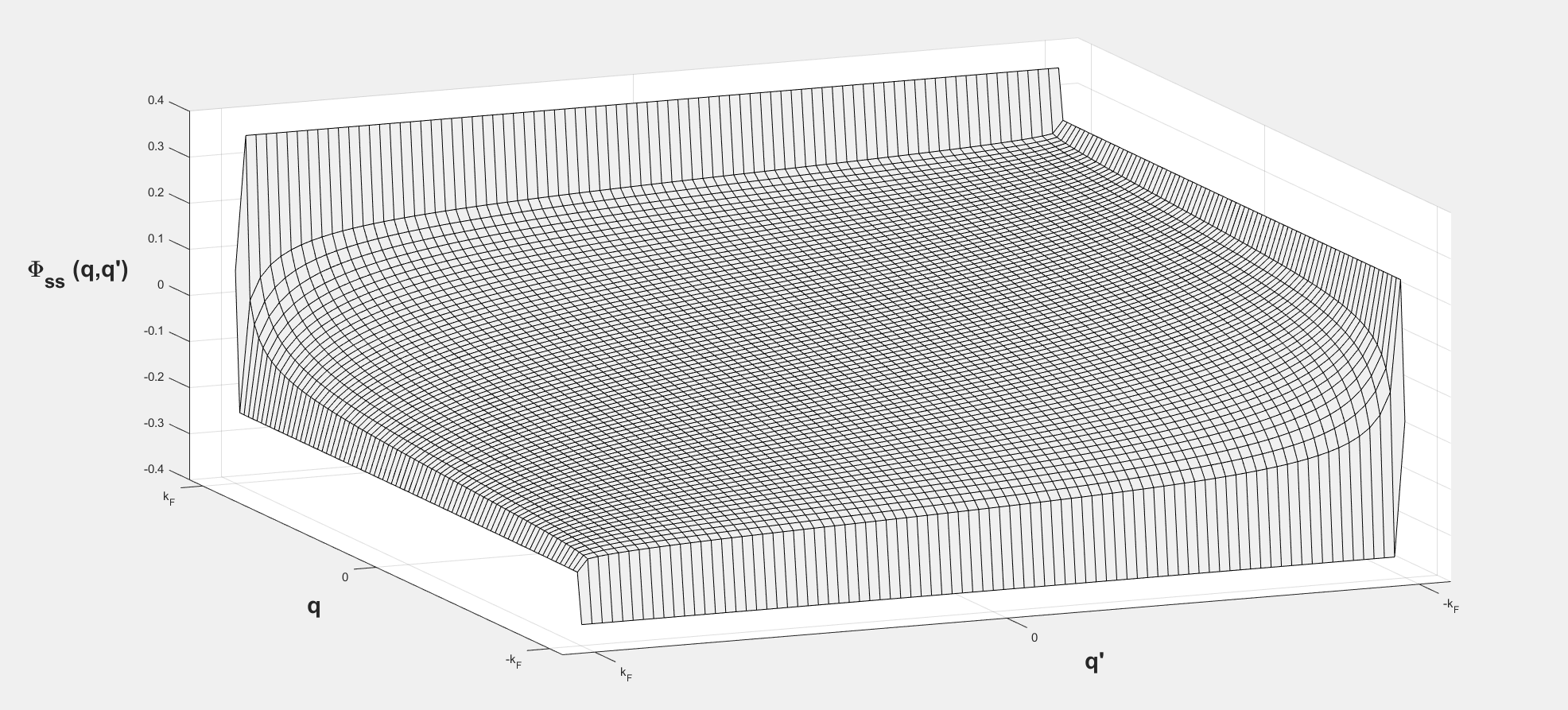}
  \centering
  \caption{The phase shift $\Phi_{ss} (q,q')$ in the $U/t \rightarrow \infty$ limit, for filling $n=0.59$. The filling $n=0.59$ is chosen from an experimental application regarding the charge transfer salt TTF-TCNQ studied in Ref. \cite{TTFTCNQ}. Note the discontinuity at the Fermi momentum $q=-k_F$ and $q=k_F$, which is a $m \rightarrow 0$ feature. In numerical applications below, we shall allow a small but vanishing magnetisation to avoid singular behaviors in the derivative of this phase shift.}
  \label{phiss}
\end{figure}

By changing variables of integration from the pseudofermion momenta $q$ and $q'$, to their rapidity counterparts $\sin k(q) / u$ and $\Lambda_s (q) / u$, where the densely filled state has limits of integration $Q = k^0 (\pi n)$ and $B = \Lambda_s^0 (\pi n_{\downarrow})$, the dressed phase shift obey (by the above construction), the following integral equations that need to be solved self-consistently:
\begin{equation}
\begin{split}
	&\Phi_{cc} (r,r') = \frac 1 {\pi} \int_{-B/u}^{B/u} \diff r'' \frac {\Phi_{sc} (r'', r')} {1+ (r-r'')^2} \\
	&\Phi_{cs} (r,r') = \frac 1 {\pi} \int_{-B/u}^{B/u} \diff r'' \frac {\Phi_{ss} (r'', r')} {1+ (r-r'')^2} - \frac 1 {\pi} \arctan (r-r') \\
     &\Phi_{sc} (r,r') = \frac 1 {\pi} \int_{-\sin Q /u}^{\sin Q /u} \diff r'' \frac {\Phi_{cc} (r'', r')} {1+ (r-r'')^2} - \frac 1 {2\pi} \int_{-B/u}^{B/u} \diff r'' \frac {\Phi_{sc} (r'', r')} {1+ (\frac{r-r''} 2)^2} - \frac 1 {\pi} \arctan (r-r') \\
	&\Phi_{ss} (r,r') = \frac 1 {\pi} \int_{-\sin Q /u}^{\sin Q /u} \diff r'' \frac {\Phi_{cs} (r'', r')} {1+ (r-r'')^2} - \frac 1 {2\pi} \int_{-B/u}^{B/u} \diff r'' \frac {\Phi_{ss} (r'', r')} {1+ (\frac{r-r''} 2)^2} + \frac 1 {\pi} \arctan (\frac {r-r''} 2) \\ 
\end{split}
\end{equation}
These phase shifts are plotted in Figs. (\ref{phicc}) - (\ref{phiss}).


\begin{figure}[h!]
  \includegraphics[width=0.99\linewidth]{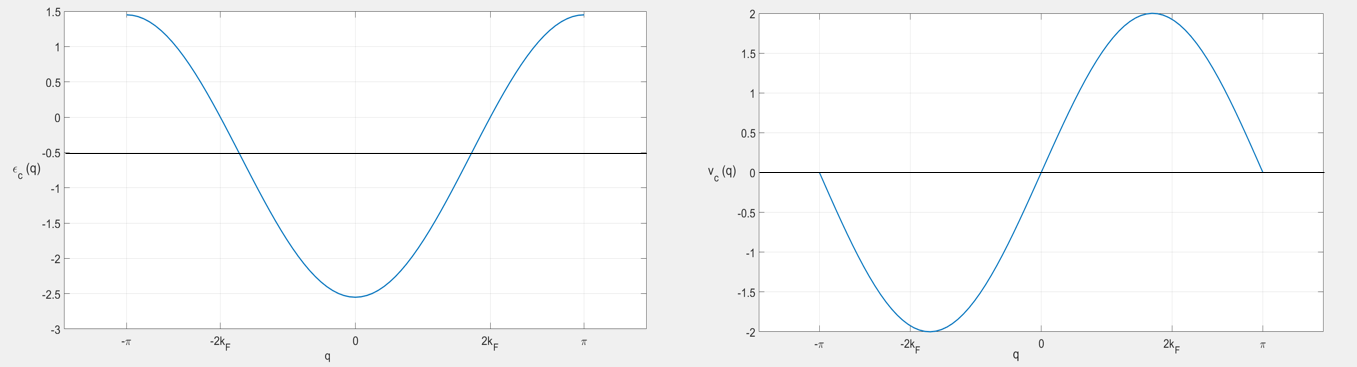}
  \centering
  \caption{The $c$ pseudofermion dispersion relation $\varepsilon_c (q)$ (right), normal ordered with $\varepsilon_c (-2k_F) = \varepsilon_c (2k_F) = 0$, where $2k_F = \pi n$ and $n=0.59$, in the $U/t \rightarrow \infty$ limit, and the corresponding group velocity of the same pseudofermion (right), revealing the even-function symmetry of the dispersion relation, with zero group velocity at $q = q_c^{\pm}$ Brillouin limits and at $q=0$. The filling $n=0.59$ is chosen from an experimental application regarding the charge transfer salt TTF-TCNQ studied in Ref. \cite{TTFTCNQ}. The corresponding dispersion relation and group velocity for the $s$ pseudofermion is not plotted in this limit due to it being completely flat and furthermore, as the magnetisation $m \rightarrow 0$, completely filled.}
  \label{epsvc}
\end{figure}

By introducing the expressions for the rapidity deviations $k^0 (q) + \Delta k(q)$ and the corresponding number occupancy deviations $N_c(q) + \Delta N_c (q)$, into the expression for $\Delta E$ of Eq. (\ref{energybands}), one can express the energy dispersion relations for the $c$ and $s$ pseudofermion in terms of these phase shifts:
\begin{equation}
\begin{split}
\varepsilon_c (q) &= -2t \cos k^0(q) + 2t \int_{-Q}^Q \diff k \sin k \,\tilde{\Phi}_{cc} (k,k^0 (q)) \\
\varepsilon_s (q) &= 2t \int_{-Q}^Q \diff k \sin k \, \tilde{\Phi}_{cs}(k,\Lambda_s^0 (q)) \\
\end{split}
\end{equation}
where $\tilde{\Phi}_{cc} (k(q),k^0 (q))$ is the resulting phase shift function after changing variables from $(q,q')$ (similarily for $\Phi_{cs}$). The dispersion relation for the $c$ pseudofermion, along with the group velocity $v_c (q) = \diff \varepsilon_c(q) /\! \diff q$ is plotted in Fig. (\ref{epsvc}). The corresponding dispersion relation for the $s$ pseudofermion reveals that it is completely flat in the $U/t \rightarrow \infty$ limit, rendering a zero group velocity for all values of pseudofermion momenta. Furthermore, as the magnetisation $m \rightarrow 0$, the $s$ band is also completely filled, with the momenta values for the Brillouin limit coinciding with those of the Fermi points, as $n_{\uparrow} = n_{\downarrow}$.
\newline

It has been shown that the pseudofermions lack any residual energy interactions \cite{pdt} \cite{pdt2}, meaning that each scattering event they undergo only makes a trivial change to an eigenstate of the system by a mere shift in the phase of the eigenstate \cite{carmeloS}. The lack of residual interactions should be understood by the $\alpha$ particle's residual interaction being cancelled by the momentum transfer term $\Phi_{\alpha} (q)$. In other words, the information recorded in the $\alpha=c,s$ particle interactions is transferred over to the momentum of the pseudofermion phase shift. The algebra needed to show this is quite lengthy and doesn't add to the physical intuition, however, the $c$ and $s$ particle's second-order interaction term is presented for example in Eq. (67) of Ref. \cite{carmelo62}, with the corresponding \textit{f-functions} appearing in Eq. (A.24) of that same reference. Intuitively, one could understand the interaction term being zero for the pseudofermions by noticing that the f-functions are proportional to either $\Phi_{\alpha} (q)$ or $(\Phi_{\alpha} (q))^2$, rendering these expressions zero if the momentum of the quantum objects being studied was $q + (2\pi/L) \Phi_{\alpha} (q)$ to begin with (hence, having zero residual shift in the interaction term, making the residual interaction term disappear).

\section{Setup for the one-electron spectral function}

\subsection{Pseudofermion operators}
According to the pseudofermion dynamical theory, electron creation and annihilation operators can be transformed unitarily into so-called rotated electron operators. We have seen that one way of representing the unitary transformation responsible for this process is to expresses the electron operators as a series of products of rotated electron operators. By only retaining the first term in the electron-rotated electron unitary transformation, we obtain an approximation to the full spectral function accounting for about 94\% of the total spectral weight in the one-electron addition case, and 98\% of the total spectral weight in the one-electron removal case, respectively \cite{carmelo94percent}. 
\newline

The $c$ and $s$ particle creation and annihilation operators can in turn be expressed in terms of the rotated electron operators. A $c$ particle creation operator $f_{c,i}^{\dag}$ creates a singly occupied site at position $i$ in the effective $c$-lattice when acting upon a rotated electron lattice site which is either doubly occupied or empty, and yielding zero otherwise. We therefore expect this operator to be of the form $\tilde{c}_{i,\uparrow}^{\dagger} (1-\tilde{n}_{i,\downarrow})+(-1)^i \tilde{c}_{i,\uparrow} \tilde{n}_{i,\downarrow}$ where the factor $(-1)^i$ guarantees the fermion algebra. An $s$ particle creation operator acts on two rotated electron lattice sites simultaneously in order to create a spin-singlet $(\uparrow,\downarrow)$ pair on a lattice position in the $s$ effective lattice, but yielding zero when acting on a state already containing such a pair. Therefore, it would be expected for this operator to be of the form $\tilde{c}_{i,\downarrow}^{\dagger} \tilde{c}_{j,\uparrow}^{\dagger} (1-\tilde{n}_{j,\downarrow})(1-\tilde{n}_{i,\uparrow}) - \tilde{c}_{i,\uparrow}^{\dagger} \tilde{c}_{j,\downarrow}^{\dagger} (1-\tilde{n}_{j,\uparrow})(1-\tilde{n}_{i,\downarrow})$ (where $i \ne j$). However, in contrast to $f_{c,i}^{\dag}$, this creation and annihilation operator pair for the $s$ particle does not obey fermion algebra, since it involves pairs of creation (or annihilation) rotated electron operators. Therefore, it needs to be adjusted by a Jordan-Wigner transformation ensuring fermion behavior \cite{JW}. Exact operator expressions of this kind is given in Ref. \cite{carmelo62}. 
\newline

The Fourier transformed version of these creation and annihilation operators are
\begin{equation}
f^{\dagger}_{\alpha,q_i} = \frac 1 {\sqrt L} \sum_j \mathrm{e}^{\mathrm{i} q_i x_j} f^{\dagger}_{\alpha,j}
\end{equation}
where the summation runs over the $c$ and $s$ effective lattice sites, respectively. However, since the pseudofermion basis offers zero residual interactions as mentioned before, which is the property we need in order to factorise the wave function into a $c$ and an $s$ part, we use another unitary transformation $\hat{W}$ which enables us to express the pseudofermion creation and annihilation operators in terms of these $\alpha = c$ or $s$ particle operators. The task that $\hat{W}$ needs to accomplish is simply to shift the momenta values of each $\alpha = c$ or $s$ particle, to the corresponding pseudofermion momenta values (which includes the phase shift discussed above). $\hat{W}$ is given in Eq. (58) of Ref. \cite{carmelo62}. As a summary, the above described procedure can be encapsulated in the following flow chart:
\newline

\noindent\fbox{\parbox{\textwidth}{Electrons $\implies$ Unitary transformation $\hat{V} \implies$ Rotated electrons (dbl. occ. is a good quantum number) $\implies \alpha =c$ or $s$ particles with momentum $q$ (densely packed GS) $\implies$ Unitary transformation $\hat{W} \implies$ pseudofermions with momentum $q + \frac {2\pi} L \Phi_{\alpha} (q)$ (for excited energy eigenstates)}}
\newline

This approach is inspired by K. Penc's study of a similar 1D Hubbard model with $U/t=\infty$ \cite{penc1}-\cite{penc4}, and expresses a factorisation of the wave function into charge and spin separated parts for all energy scales. 
To understand how this is done using the PDT, let us first define the one electron spectral functions in question in the Lehman representation:
\begin{equation}
\begin{split}
A^+ (k,\omega) &= \sum_{f_{N \! + \! 1},\sigma} \big\lvert \langle f_{N+1} | c_{k,\sigma}^{\dag} | GS \rangle \big\rvert^2 \delta (\omega - E_{f_{N+1}}) \\
A^- (k,\omega) &= \sum_{f_{N \! - \! 1},\sigma} \big\lvert \langle f_{N-1} | c_{k,\sigma} | GS \rangle \big\rvert^2 \delta (\omega - E_{f_{N-1}}) \\
\end{split}
\end{equation}
for the one electron creation, $A^+ (k,\omega)$, and annihilation, $A^- (k,\omega)$, spectral functions. Here, the summation $f_{N\pm 1}$ goes over all excited energy eigenstates with $N\pm 1$ electrons, while $GS$ stands for the $N$-electron ground state of the system. $c_{k,\sigma}^{\dag}$ is the Fourier transform of the electron creation operator $c_{i,\sigma}^{\dag}$ acting on the physical lattice, where $i=1,2,\ldots,N_a$. As stated above, nearly all spectral weight is retained by approximating $c_{k,\sigma}^{\dag}$ with its rotated electron counterpart $\tilde{c}_{k,\sigma}^{\dag}$ (and similarily for the annihilation operator $c_{k,\sigma}$). The pseudofermion phase shift are generally non-zero, meaning that the ground state and the final states will have different boundary conditions for each pseudofermion. This implies that the evaluation of the matrix overlaps leads to Anderson's orthogonality catastrophe in the thermodynamic limit \cite{orthcat1} \cite{orthcat2}. We should therefore expect the spectral weight of the low-lying excitations to decay exponentially. Moreover, since the phase shift is state dependent, we would expect a different decay exponent for each ground state $\rightarrow$ final state transition, where the final state is characterised by $c$ and $s$ pseudofermion dispersing throughout their energy bands as given by Eq. (\ref{energybands}). 
\newline

Before we proceed with evaluating the matrix overlaps of the spectral functions, a brief comment on the choice of spin projection is in order, as it turns out there are specific choices for $\sigma$ in the one electron addition and removal cases for the $m \rightarrow 0$ limit. We remind ourselves that within a given final state, there are additionally $2S_{\alpha}$ number of non-dynamical states, reachable by successive application of the spin raising operator onto said final state (sometimes referred to as "building a tower of states", in this case from $-S_{\alpha}$ to $S_{\alpha}$, for both $\alpha=c$ and $s$). Now, a typical final state with spin projection value $-S_{s}+1$ can be written as $| - S_{s}+1 \rangle \sim \hat{S}^{\dagger}_s | -\!S_{s} \rangle$ (up to a normalisation constant). A matrix overlap between such a final state and a ground state would then be $\langle -S_{s}+1 | \tilde{c}_{i,\sigma} | GS \rangle \sim \langle -S_{s} | \hat{S}_s \tilde{c}_{i,\sigma} | GS \rangle = - \delta_{\sigma,\uparrow} \sigma \langle -S_{s} | \hat{S}_s \tilde{c}_{i,-\sigma} | GS \rangle$, by use of the commutator $[ \hat{S}_s , \tilde{c}_{i,\sigma}] = - \delta_{\sigma,\uparrow}\sigma \tilde{c}_{i,-\sigma}$ and the fact that $\hat{S}_s |GS\rangle = 0$. This means that by choosing, in case of the one electron removal spectral function, $\tilde{c}_{i,\downarrow}(1-\tilde{n}_{i,\uparrow})$, all final states will indeed be lowest weight states. For the one electron creation spectral function, the corresponding choice turns out to be $\tilde{c}_{i,\uparrow}^{\dagger}(1-\tilde{n}_{i,\downarrow})$.
\newline

The $\alpha=c,s$ particles have momentum that we label $q_{\alpha,i}$ or just $q_i$ depending on the context. The corresponding pseudofermion momentum will then be $\bar{q}_{\alpha,i}$ (or just $\bar{q}_{i}$) according to:
\begin{equation}
\begin{split}
\left.
\begin{alignedat}{4}
	&\bar{q}_i  = q_i + \frac {2 \pi} L \Phi_{\alpha} (q_i)  \\
        &q_i  = \frac {2 \pi} L (i-1) - \frac {2 \pi} L \Delta N_{F\alpha}^{-}- \frac {\pi} L N_{\alpha} + \frac {\pi} L   \\
	\\
	&\Phi_{\alpha}(q_i) =\phi_0 + \sum_{\alpha'=c,s} \sum_{q'} \Phi_{\alpha,\alpha'} (q_i, q') \Delta N_{\alpha'} (q') \quad \\	     
\end{alignedat}
\right\}
\qquad i = 1,2,\dots,N_{\alpha}+\Delta N_{F\alpha}^+ + \Delta N_{F\alpha}^- \\
\end{split}
\end{equation}
where $\Delta N_{F\alpha}^{\pm}$ is the number of $\alpha$ pseudofermions created ($>0$) or removed ($<0$) at the $\alpha$ positive or negative Fermi point (with momentum equal to $q_{F\alpha}^-$ or $q_{F\alpha}^+$, respectively). The fact that these numebrs are included here is due to a peculiarity of the thermodynamic limit: adding and removing pseudofermions at the Fermi level is a zero-energy process, in contrast to a model where the thermodynamic limit is not taken. We therefore allow for a creation and/or annihilation of a finite number of pseudofermions at the Fermi level (on both sides) when we generate the ground state going into the spectral function calculation. Such a ground state will be called $\lvert \widetilde{GS} \rangle$. Notice that the usual $N$ particle system is obtained when $\Delta N_{F\alpha}^{+}= \Delta N_{F\alpha}^{-} = 0$ for the "true" ground state of the system $\lvert GS \rangle$. These numbers are important since they define the pseudofermion sea, inside of which is a continuum of occupied momenta values, outside of which are unoccupied momenta values. This renders the effective Fermi momenta to be $\kappa_{\alpha}^{\pm} = q_{F\alpha}^{\pm} + \Delta q_{F\alpha}^{\pm}$, where $\Delta q_{F\alpha}^{\pm} = \pm (2\pi / L) \Delta N_{F\alpha}^{\pm}$. $\phi_0$ is the fixed momentum shift that might occur due to the change in the type of quantum numbers as described by Takahashi's equations (going from either integer to half-odd integer or vice versa), equal to $-1/2,0$ or $+1/2$.  

Since the ground state is a densely packed state in terms of the $c$ and the $s$ particles, it can be written as the operator product over all occupied momenta values:
\begin{equation}
\lvert \widetilde{GS} \rangle = \prod_{\alpha} \prod_{q=\kappa_{\alpha}^{-}}^{\kappa_{\alpha}^{+}} f_{\alpha,q}^{\dagger} | 0 \rangle \label{gs}
\end{equation}

\subsection{Excited state characterisation}
In the case of the one electron removal spectral function, as stated above, the rotated electron operator is $\tilde{c}_{i,\downarrow} (1-\tilde{n}_{i,\uparrow})$. Since one unit of charge is removed on a singly occupied site, this means that a $c$ particle is removed. Also, an $s$ particle is removed since in the lowest weight state constituting the ground state, all $\downarrow$-spins are paired with $\uparrow$-spins. By removing one $\downarrow$-spin, one spin singlet $(\uparrow,\downarrow)$ pair is broken up. Furthermore, the quantum numbers describing the occupancies in the $c$ band change from being half-odd integers to integers, inducing a global momentum shift equal to $\pm N \cdot (\pi/L) = \pm \pi n$. In the case of the one electron addition spectral function, clearly one unit of charge is added on a singly occupied site, creating one $c$ particle. However, since we are adding a $\uparrow$-spin, we are not creating a new $s$ particle. Instead, the number of possible $\uparrow$-spins that the $\downarrow$-spins in the system can form spin-singlet pairs with, have increased by one. Hence, we have increased an $s$ particle hole, corresponding to the extra available slot in the $s$ band. These two processes allocate spectral weight on the $(k,\omega)$ plane according to (for clarity, in the following we skip the (1/L) correction terms):

\begin{equation}
\begin{split}
\begin{alignedat}{4}
	&\text{Removal} \\
	&\omega  = \varepsilon_c (q_c) + \varepsilon_s (q_s) \qquad \\
     &k = q_c + q_s - 2 \pi n \, \phi_0 \qquad \\	  
	&q_c \in [-\pi n , \pi n] \\
	&q_s \in [-\pi n_{\downarrow} , \pi n_{\downarrow}] \\
\end{alignedat}
\begin{alignedat}{4}
	&\text{Addition} \\
	&\omega  = \varepsilon_c (q_c) - \varepsilon_s (q_s) \\
     &k = q_c - q_s  \\     
	&q_c \in [-\pi , - \pi n] \cup [ \pi n , \pi] \\
	&q_s \in [-\pi n_{\downarrow} , \pi n_{\downarrow}] \\ 
\end{alignedat}
\end{split} \label{setup}
\end{equation}
where we note that the energy dispersions are normal ordered such that $\varepsilon_{\alpha} (q_{F\alpha})=0$, and where $\varepsilon_c (q)$ is depicted in Fig. (\ref{epsvc}). Note that in our $U/t \rightarrow \infty$ calculation of the removal spectral function below, $\varepsilon_s (q) = 0$ for all $q$, with a vanishing number of unoccupied momenta values above the Fermi points as $m \rightarrow 0$. Now, in both the one electron addition and removal spectral function cases, the rotated electron operator within the matrix overlap creates and/or destroys a small amount of pseudofermions, thus exciting the ground state. Let us denote the operator responsible for this action by $\hat{\Theta} (k)$. By acting on the final state, however, we obtain a densely packed pseudofermion occupancy configuration by being acted upon by the hermitian conjugate of this operator $\hat{\Theta}^{\dagger} (k)$ (equivalent of letting $\hat{\Theta} (k)$ acting on the bra instead of the ket). This procedure is formalised in Refs. \cite{carmelo62} and \cite{carmelo64}, and can intuitively be understood as an $\alpha$ pseudofermion being created at momentum position $q_{\alpha}$ as an active scattering center upon the existing sea of occupied momenta positions. Even though each of these occupied positions registers a phase shift $\Phi_{\alpha} (q_{\alpha})$ in this process, since the sea in the thermodynamic limit is compact, it is only the shifts at the effective Fermi boundaries which will be of dynamical importance, i.e. $\Phi_{\alpha} (q_{F\alpha})$.

A typical final state can thus be written as
\begin{equation}
\hat{\Theta}^{\dagger} (k) | f \rangle = \prod_{\alpha} \prod_{\bar{q}=\bar{\kappa}_{\alpha}^{-}}^{\bar{\kappa}_{\alpha}^{+}} f_{\alpha,\bar{q}}^{\dagger} | 0 \rangle \label{fs}
\end{equation}
where $\bar{q} = q + (2\pi / L)\Phi_{\alpha} (q)$ and $\bar{\kappa}^{\pm}_{\alpha} = \kappa^{\pm}_{\alpha} + (2\pi / L)\Phi_{\alpha} (q_{F\alpha}^{\pm})$.

\subsection{Slater determinant}
To evaluate the matrix overlap between the states given in Eqs. (\ref{gs}) and (\ref{fs}), we need to evaluate a Slater determinant consisting of all the anticommutators between the unshifted $c$ and $s$ particles in the ground state, and the corresponding shifted pseudofermion final state. We remember that the eigenstates factorise completely, by virtue of having zero residual energy interactions, meaning that the following mathematical treatment applies equally to $\alpha=c$ and $s$. By using the fermion anticommutation relations in the real lattice,
\begin{equation}
\{ f^{\dagger}_{\alpha,\bar{q}},f_{\alpha',q'} \} = \frac {\delta_{\alpha,\alpha'}} L \sum_{j,j'} \mathrm{e}^{\mathrm{i} (\bar{q}j-q'j')} \{ f^{\dagger}_{\alpha,j},f_{\alpha',j'} \}
\end{equation}
we arrive at, after using periodic boundary conditions to evaluate the sum,
\begin{equation}
\{ f^{\dagger}_{\alpha,\bar{q}},f_{\alpha',q'} \} = \frac {\delta_{\alpha,\alpha'}} L \mathrm{e}^{\mathrm{i} (\bar{q}-q')/2} \mathrm{e}^{\mathrm{i} \pi \Phi_{\alpha} (q)} \frac {\sin (\pi \Phi_{\alpha}(q))} {\sin (\bar{q}-q')/2)}
\end{equation}
where we notice that the ground state momenta $q'$ are have no phase shift.
\newline

The quantity to evaluate hence becomes (where $M_{\alpha}=N_{\alpha}+\Delta N_{F\alpha}^+ + \Delta N_{F\alpha}^-$ denotes the total number of pseudofermions in the excited energy eigenstate):
\begin{equation}
\begin{split}
\Big\lvert \langle f | \hat{\Theta} (k) \lvert \widetilde{GS} \rangle \Big\rvert^2 &= \Big\lvert \det(\{ f^{\dagger}_{\alpha,\bar{q}_i},f_{\alpha,q'_j} \}_{1 \le i,j \le M}) \Big\rvert^2 = \\
& = \frac 1 {L^{2M}} \prod_{i=1}^M \sin^2 \big( \pi \Phi(q_i) \big) \cdot \prod_{i=1}^{M-1} \bigg( \sin^2 \frac {\pi i} L \bigg)^{M-i} \cdot \prod_{i=1}^{M-1} \prod_{j=i+1}^{M} \bigg( \sin^2 \frac {\bar{q}_i-\bar{q}_j} 2 \bigg) \cdot \prod_{i=1}^{M} \prod_{j=1}^{M} \bigg( \sin^{-2} \frac {\bar{q}_i-q_j} 2 \bigg) \label{slater}
\end{split}
\end{equation}
This expression looks cumbersome, but there are some simplifications to make in the thermodynamic limit, namely by realising that,
\begin{equation}
\begin{split}
\prod_{i=1}^{M} \prod_{j=1}^{M} \bigg( \sin^{2} \frac {\bar{q}_i-q_j} 2 \bigg) = \prod_{i=1}^{M-1} \prod_{j=i+1}^{M} \bigg( \sin^2 \frac {\bar{q}_i-q_j} 2 \bigg) \cdot \prod_{i=1}^{M} \bigg( \sin^2 \frac {\bar{q}_i-q_i} 2 \bigg) \cdot \prod_{i=1}^{M-1} \prod_{j=i+1}^{M} \bigg( \sin^2 \frac {q_i-\bar{q}_j} 2 \bigg)
\end{split}
\end{equation}
where in the thermodynamic limit (and using shorthand $\Phi_i$ for $\Phi_{\alpha} (q_i)$ where it is implicitly understood that in the following, $\alpha$ is fixed to be either $c$ or $s$),
\begin{equation}
\begin{split}
\prod_{i=1}^{M} \bigg( \sin^2 \frac {\bar{q}_i-q_i} 2 \bigg) = \prod_{i=1}^{M} \bigg( \sin^2 \frac {\pi \Phi_i} L \bigg) \approx  \prod_{i=1}^{M} \bigg( \frac {\pi \Phi_i} L \bigg)^2 = L^{-2M} \prod_{i=1}^M \bigg( \pi \Phi_i \bigg)^2
\end{split}
\end{equation}
cancelling out the pre-factor $L^{2M}$ in Eq. (\ref{slater}). Therefore,  Eq. (\ref{slater}) simplifies to
\begin{equation}
\begin{split}
& \Big\lvert \langle f | \hat{\Theta} (k)\lvert \widetilde{GS} \rangle \Big\rvert^2 = \\ 
& \prod_{i=1}^M \bigg( \frac {\sin \big( \pi \Phi_i \big)} {\pi \Phi_i} \bigg)^2  \cdot \prod_{i=1}^{M} \bigg( \sin^2 \frac {\pi i} L \bigg)^{M-i} \cdot \prod_{i=1}^{M-1} \prod_{j=i+1}^{M} \bigg( \sin^2 \frac {\bar{q}_i-\bar{q}_j} 2 \bigg) \cdot \prod_{i=1}^{M-1} \prod_{j=i+1}^M \bigg( \sin^{-2} \frac {\bar{q}_i-q_j} 2  \sin^{-2} \frac {q_i-\bar{q}_j} 2 \bigg) \label{slater2}
\end{split}
\end{equation}
In the following section, we shall refer to this expression as the \textit{lowest peak weight} $A^{(0,0)}$ which is a functional depending on all phase shift of the occupied pseudofermion momenta within the excited energy eigenstate: \newline
$A^{(0,0)} = A \big( \big\{ \Phi(q_1), \Phi(q_2), \dots , \Phi(q_M) \big\} \big)$.

\section{Lowest peak weight $A^{(0,0)}$: General approach}

\subsection{Initial inspiration}
In order to obtain a closed form expression for the lowest peak weight $A^{(0,0)}$, we would need to evaluate Eq. (\ref{slater2}). For this purpose, even though this expression is complete, a brute-force approach would be problematic due to numerical difficulties of managing all pseudofermion momenta values (where all phase shift are state dependent) in the thermodynamic limit. It would also be misleading to evaluate this product using "smaller" systems due to $(1/L)$ approximations already made in the derivation of it. However, the evaluation of this matrix overlap for both $\alpha$ pseudofermions could in theory be used to reproduce known results for the spectral weight in the entire $(k,\omega)$ plane. We will particularly compare our results with previously obtained results for the one electron removal spectral function in the $U/t=\infty$ limit \cite{penc1} - \cite{penc4}. In those references, closed form expressions were obtained for smaller system sizes where the momentum $Q$ of the spin part, affected the charge part via a twisted boundary condition $\mathrm{e}^{\mathrm{i} Q}$ to guarantee periodicity, similar to the phase shifts we encounter in the PDT. The dynamical  quantum objects in those references were spinless fermions on the one hand and chargeless spins on the other, where the spin part mapped onto a 1D antiferromagnetic Heisenberg spin hamiltonian \cite{Uinfty1}. In our approach, a similar charge and spin separation is obtained and used to factorise all eigenstates, by use of the $c$ and $s$ pseudofermions.
\newline

To obtain closed forms expressions of the spectral functions in Refs. \cite{penc1} - \cite{penc4}, the recurrence relation $A_{Q+2\pi} A_{Q-2\pi} \big/ (A_Q)^2$, where $A_Q$ is the expression for the Slater determinant valid for a constant phase shift $Q$, was evaluated. After cancelling out of common factors, solving the recurrence relation was a matter of algebra. In the following, we will be inspired by this technique to try to attain a similar "simple" expression for the spectral function in the general $U/t$ case. Therefore, using our notation, we will apply a similar recurrence relation as used in Refs. \cite{penc1} - \cite{penc4}, namely $\Phi_{\alpha} (q) \rightarrow \Phi_{\alpha} (q) \pm 1$, and apply these shifts to the entire pseudofermion sea of occupied states. In other words, we will evaluate the expression
\begin{equation}
\frac {A \big( \big\{ \Phi(q_1)+1, \Phi(q_2)+1, \dots , \Phi(q_M)+1 \big\} \big) A \big( \big\{ \Phi(q_1)-1, \Phi(q_2)-1, \dots , \Phi(q_M)-1 \big\} \big)} {\big( A \big( \big\{ \Phi(q_1), \Phi(q_2), \dots , \Phi(q_M \big\} \big)^2}, \label{phipm1}
\end{equation}
which means that we will treat the state dependent phase shifts $\Phi (q)$ as independent parameters.
\newline

\subsection{Setup of the recurrence relation}
Imagine now that we rewrite Eq. (\ref{slater2}) twice; once with $\Phi (q_i) \rightarrow \Phi (q_i) + 1$ and once with $\Phi (q_i) \rightarrow \Phi (q_i) - 1$, and then plug those into Eq. (\ref{phipm1}). The majority of factors cancel out in the resulting expression. For example $\sin^2 \big( \pi [\Phi(q_i) \pm 1] \big) = \sin^2 \big( \pi \Phi(q_i) \big)$, and all the factors containing $\bar{q}_i-\bar{q}_j$ cancel, since $\bar{q}_i-\bar{q}_j = q_i-q_j + \big[ 2\pi / L \big] \big( \Phi (q_i) - \Phi(q_j) \big)$, which is insensitive to global shifts of $+1$ or $-1$:  $\Phi (q_i) \pm 1 - \big( \Phi(q_j) \pm 1 \big)= \Phi (q_i) - \Phi(q_j)$. 
\newline

The only product which does not cancel trivially is the last one:
\newline
\begin{equation}
\prod_{i=1}^{M-1} \prod_{j=i+1}^{M} \bigg( \sin^{-2} \frac {\bar{q}_i-q_j} 2 \bigg) \bigg( \sin^{-2} \frac {q_i-\bar{q}_j} 2 \bigg). \label{shifts}
\end{equation}

\begin{figure}[h!]
  \includegraphics[width=\linewidth]{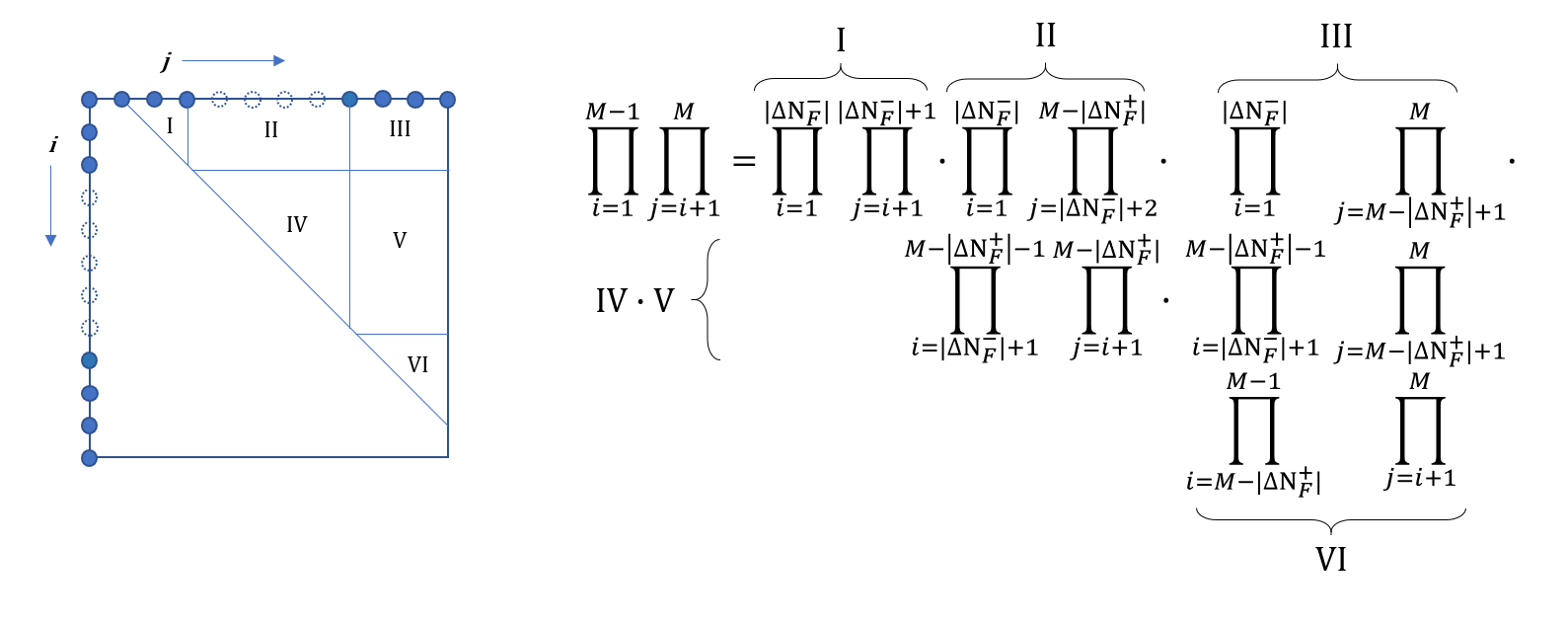}
  \caption{A schematic illustrating the productorials given in Eq. (\ref{shifts}), where the two indices $i$ and $j$ run vertically and horisontally, respectively, over all $M$ pseudofermions in the product. Notice that the products II, IV, and V, have at least one of their indices running across the thermodynamic bulk of the pseudofermions (for product IV both indices run over the bulk), while for the products I, III  and VI, none of the indices run over the bulk and are confined to the few pseudofermions located at the left or right Fermi points, respectively. Consequently, productorials I, III and VI will in most cases be trivial to write down exactly, while it would in general be impossible to find an exact expression for productorials II, IV and V, owing to the sheer number of factors in those expressions.}
  \label{proddomain}
\end{figure}

In order to deal with this expression, we divide the product domain (likened to a triangle with width and height $M-1$) into several smaller subdomains, as schematically shown in figure \ref{proddomain}. We will also use the fact that we are only interested in the thermodynamic limit, which renders individual phase shifts inside the occupied sea "uninteresting" as momentum values form a continuum there. On the other hand, the phase shifts on the edges of the occupied sea, $\Phi(\iota q_F)$ (where $\iota=\pm 1$), \emph{define} this sea and are hence explicitly needed. Given that these phase shifts will play a crucial role in the derivations to come, we will introduce a shorthand notation for them: $\Phi(\iota q_F)=\phi_{\iota}$. To facilitate this calculation, we approximate $\bar{q}_{i+1} - \bar{q}_i \approx 2 \pi / L$ over neighbouring pairs of momenta, which leads to a pairwise cancellation of factors whenever $i$ lies within the occupied sea (this approxiation is on the order of $(1/L)^2$).  This supports the notion that, in the thermodynamic limit, we should expect the spectral function to only depend on the phase shift at the edges of the occupied sea, $\lim_{L\rightarrow \infty} A \big( \big\{ \Phi(q_1), \Phi(q_2), \dots , \Phi(q_M \big\} \big) = A(\phi_-, \phi_+)$. We will furthermore approximate the values of all phase shift associated with pseudofermions created or annihilated at the edges of the occupied sea, with the value of the phase shift \emph{at} the edge: $\Phi(q_i) \approx \Phi(-q_F)$ for $i = 1, \dots, \Delta N_F^{-}$, and $\Phi(q_i) \approx \Phi(q_F)$ for $i =M - \Delta N_F^{+} + 1, \dots, M$. This is a safe approximation also on the order of $(1/L)^2$, given that the numbers $\lvert \Delta N_F^{\iota} \rvert$ are microscopic when compared to $M \rightarrow \infty$ (in this investigation, $\lvert \Delta N_F^{\iota} \rvert$ is either $0$ or $1$).

Calculating the products region by region, as depicted in Fig. (\ref{proddomain}), we notice that it is only for region IV where both momenta indices run over the thermodynamic bulk of the pseudofermions, having none of the phase shifts equal to those at a Fermi point. Rather, in this domain, they depend on the value of the pseudofermion momentum $q$, where $q_F^-<q<q_F^+$. In contrast, for region I, $\Phi (q_i) = \Phi (q_j) = \phi_-$, while oppositely in region VI,  $\Phi (q_i) = \Phi (q_j) = \phi_+$. For region II we have that $\Phi (q_i) =  \phi_-$ but $q_j \in (q_F^- , q_F^+´)$ in the expression for $\Phi (q_j)$, and similarily for region V, where $\Phi_j =  \phi_+$ but $q_i \in (q_F^- , q_F^+´)$ in the expression for $\Phi (q_i)$. As before, the index $i$ or $j$ only serves a purpose in the discrete system (i.e. the quantization in terms of the momentum spacing $2 \pi / L$), which smears out to a continuum when $L \rightarrow \infty$. In this limit, $\Phi(q_i) \rightarrow \Phi (q)$, where $q$ without indices denotes a continuous variable.
\newline

\subsection{Evaluation of the recurrence relation in the thermodynamic limit}
We will now show how to evaluate these products in the thermodynamic limit, and simplify the result when applying the recursive relation described by Eq. (\ref{phipm1}). The productorials in the regions I and VI are quite similar to each other, so we will only show the result for region VI. The evaluation of the productorial in region I can then be obtained by replacing $\phi_+$ with $\phi_-$ and $\lvert \Delta N_F^{+} \rvert$ by $\lvert \Delta N_F^{-} \rvert$, respectively. By letting the productorial in region VI be denoted by $A_{VI} (\phi_+)$, we notice first that
\begin{equation}
\frac {\bar{q}_i - q_j} 2 = \frac {\pi} L (i-j) +  \frac {\pi} L \Phi_i 
\end{equation}

which means that in region VI, the productorial becomes
\begin{equation}
\begin{split}
A_{VI} (\phi_+) = \prod_{i=M-\lvert \Delta N_F^{+} \rvert}^{M-1} \prod_{j=i+1}^{M} & \bigg( \sin^{-2}  \frac {\pi} L (i-j) +  \frac {\pi} L \phi_+  \bigg) \bigg( \sin^{-2}  \frac {\pi} L (i-j) -  \frac {\pi} L \phi_+  \bigg) =  \\
& \\
& = \prod_{i=i}^{\lvert \Delta N_F^{+} \rvert} \Bigg[ \sin \bigg( \frac {\pi i} L -  \frac {\pi} L \phi_+ \bigg)  \sin \bigg( \frac {\pi i} L +  \frac {\pi} L \phi_+ \bigg) \Bigg]^{2\big( i-\lvert \Delta N_F^{+} \rvert-1 \big)} =\\
& \\
& =  \bigg( \frac{\pi} L \bigg)^{-2\lvert \Delta N_F^{+} \rvert \big( \lvert \Delta N_F^{+} \rvert +1 \big)} \cdot \prod_{i=i}^{\lvert \Delta N_F^{+} \rvert} \bigg(i^2-\phi_+^2 \bigg)^{2 \big( i-\lvert \Delta N_F^{+} \rvert-1 \big)} 
\end{split}
\end{equation}

In order to evaluate the recursive expression given in Eq. (\ref{phipm1}), we need to evaluate this expression with the corresponding "phase shift shifts" (in this case by letting $\phi_+ \rightarrow \phi_+ \pm1$).
\newline

Following the setup of Eq. (\ref{phipm1}), we obtain, after some simplifications, that
\begin{equation}
\begin{split}
\frac {A_{VI} (\phi_+ + 1) A_{VI} (\phi_+ -1)} {\big( A_{VI} (\phi_+) \big)^2} = \Bigg[ \bigg( \frac {\phi_+^2 - 1} {\phi_+^2}\bigg)^{\lvert \Delta N_F^{+} \rvert} \cdot \frac { \phi_+^2-1 } { \phi_+^2- \big( \lvert \Delta N_F^{+} \rvert +1 \big)^2}  \Bigg]^2 \label{Asix}
\end{split}
\end{equation}
and similarily for region I, i.e. for $A_{I} (\phi_-)$, by replacing $\phi_+$ by  $\phi_-$ and $\lvert \Delta N_F^{+} \rvert$ by $\lvert \Delta N_F^{-} \rvert$, respectively.
\newline

After some algebraic manipulation, one can see that the thermodynamic limit of the productorial for region III is constituted entirely of factors of the form $\sin \big( \pi n +\mathcal{O} (1  / L)\big)$. This means that in the recursion expression, i.e. the corresponding expression to Eq. (\ref{Asix}) but for $A_{III}$,  we obtain unity. This might not be surprising given that in this region, we are evaluating the overlap between a ground state and an excited energy eigenstate with pseudofermion occupations at the opposite end of their respective occupied seas.
\newline

For regions II, V and IV, we obtain lengthier and less obvious cancellations of factors, together with products we can not find closed expressions for, without doing some further assumptions as we shall see below. The main idea when taking the thermodynamic limit of these products, is that we are only interested in factors of the form $\sin ( \alpha_1 / L )$, where $\alpha_1$ is a finite number. These factors approximate to $\alpha_1 / L$, which means that in the case of $\alpha_1$ containing any of the phase shifts, it will be sensitive to the $\pm 1$ shifts applied earlier and hence contribute to the expression given in Eq. (\ref{phipm1}). Any other factors, either of the same form  $\sin ( \alpha_1 / L )$ with $\alpha_1$ not containing a phase shift, or of the form $\sin (\alpha_0 +  \alpha_1 / L ) \rightarrow \sin(\alpha_0)$ when $L \rightarrow \infty$, will cancel out in Eq. (\ref{phipm1}). \label{limitproc}
\newline

To get a flavour of this calculation, we can investigate the product II as an example. Since we assume $\Phi (q_i) = \phi_-$ whenever $i=1, \dots , \lvert \Delta N_F^{-} \rvert$, we will always have a fringe of factors that are assumed to have the same phase shift. This will lead to some simplification in the evaluation of the productorials:
\begin{equation}
\begin{split}
A_{II} (\phi_-) & = \prod_{i=1}^{M} \prod_{j=\lvert \Delta N_F^{-} \rvert + 2}^{M-\lvert \Delta N_F^{+} \rvert}  \sin^{-2} \bigg( \frac {\pi} L (i-j)+  \frac {\pi} L \phi_- \bigg)  \sin^{-2} \bigg( \frac {\pi} L (i-j) -   \frac {\pi} L \Phi_j \bigg) = \\
& = \prod_{i=1}^{M-\lvert \Delta N_F^{+} \rvert-2 \big( \lvert \Delta N_F^{-} \rvert+1 \big)} \Bigg[ \sin^{-2} \bigg( \frac {\pi i} L - \frac {\pi} L \big( M-\lvert \Delta N_F^{+} \rvert-\lvert \Delta N_F^{-} \rvert \big)+ \frac {\pi} L\phi_- \bigg) \Bigg]^{\lvert \Delta N_F^{-} \rvert} \cdot \\ 
& \cdot \prod_{i=1}^{\lvert \Delta N_F^{-} \rvert} \Bigg[  \sin^{-2} \bigg( \frac {\pi i} L + \frac {\pi} L -  \frac {\pi} L \phi_- \bigg) \sin^{-2} \bigg( \frac {\pi i} L - \frac {\pi} L \big( M-\lvert \Delta N_F^{+} \rvert \big) +  \frac {\pi} L \phi_-  \bigg)\Bigg]^i  \cdot \\
& \cdot \prod_{i=1}^{\lvert \Delta N_F^{-} \rvert} \prod_{j=\lvert \Delta N_F^{-} \rvert + 2}^{M-\lvert \Delta N_F^{+} \rvert} \sin^{-2} \bigg( \frac {\pi} L ( i-j) - \frac {\pi} L \Phi_j \bigg) \label{prodII}
\end{split}
\end{equation}

The next step is to multiply this expression where all the phase shift have been shifted by $+1$, with the same expression where all the  phase shift have been shifted by $-1$, and then divided by the unshifted expression squared, following the prescription given in Eq. (\ref{phipm1}). This gives, for the first productorial of Eq. (\ref{prodII}),
\begin{equation}
\begin{split}
& \left[ \frac {\sin^{-2} \bigg( \frac {\pi} L\big( M - \lvert \Delta N_F^{+} \rvert-\lvert \Delta N_F^{-}\rvert \big) - \frac {\pi} L \phi_- \bigg) \sin^{-2} \bigg( \frac {\pi} L\big( \lvert \Delta N_F^{-} \rvert + 1 \big) - \frac {\pi} L \phi_- \bigg)} {\sin^{-2} \bigg( \frac {\pi} L\big( M - \lvert \Delta N_F^{+} \rvert-\lvert \Delta N_F^{-}\rvert -1 \big) - \frac {\pi} L \phi_- \bigg) \sin^{-2} \bigg( \frac {\pi} L\big( \lvert \Delta N_F^{-} \rvert + 2 \big) - \frac {\pi} L \phi_- \bigg)}\right]^{\lvert \Delta N_F^{-} \rvert} = \\
& = \Bigg(\frac {\lvert \Delta N_F^{-} \rvert + 1 - \phi_-} {\lvert \Delta N_F^{-} \rvert + 2 - \phi_-} \Bigg)^{-2 \lvert \Delta N_F^{-} \rvert} + \mathcal{O}(1/L) \label{prodIIpart1}
\end{split}
\end{equation}

Using the same treatment, the second product in Eq. (\ref{prodII}) only contributes with its first factor, given that the second factor gives a constant (i.e. phase shift independent) zeroth order expression, which then cancels out entirely in the recursion formula. The first product however, indeed does have a thermodynamic expansion involving the phase shift:
\begin{equation}
\prod_{i=1}^{\lvert \Delta N_F^{-} \rvert} \Bigg[  \sin^{-2} \bigg( \frac {\pi i} L + \frac {\pi} L -  \frac {\pi} L \phi_- \bigg) \Bigg]^i \rightarrow  \prod_{i=1}^{\lvert \Delta N_F^{-} \rvert} \bigg( \frac {\pi} L \bigg)^{-2i} \bigg( i+1-\phi_- \bigg)^{-2i}, \qquad L \rightarrow \infty.
\end{equation}

Applying the recursion prescription to this expression, we obtain
\begin{equation}
\left[ \frac {(1-\phi_-)\big( \lvert \Delta N_F^{-} \rvert + 2 - \phi_- \big)^{\lvert \Delta N_F^{-} \rvert}} {\big( \lvert \Delta N_F^{-} \rvert + 1 - \phi_-\big)^{\lvert \Delta N_F^{-} \rvert+1}} \right]^{-2}. \label{prodIIpart2}
\end{equation}

Now, multiplying Eqs. (\ref{prodIIpart1}) and (\ref{prodIIpart2}) together, we see that some factors cancel out, with only $(1-\phi_-)^{-2} \cdot$ \newline $\cdot ( \lvert \Delta N_F^{-} \rvert + 1 -\phi_-)^{2}$ surviving. The last productorial of Eq.(\ref{prodII}), simplifies under the same treatment to
\begin{equation}
\prod_{i=\lvert \Delta N_F^{-} \rvert + 2}^{M-\lvert \Delta N_F^{+} \rvert} \: \frac {\sin^{-2} \bigg( \frac {\pi i} L+ \frac {\pi} L \Phi_i \bigg) \sin^{-2} \bigg( \frac {\pi i} L - \frac {\pi} L\big( \lvert \Delta N_F^{-} \rvert + 1 \big) + \frac {\pi} L \Phi_i \bigg)} {\sin^{-2} \bigg( \frac {\pi i} L+ \frac {\pi} L \Phi_i - \frac{\pi} L \bigg) \sin^{-2} \bigg( \frac {\pi i} L - \frac {\pi} L \lvert \Delta N_F^{-} \rvert + \frac {\pi} L \Phi_i \bigg)} \label{prodIIpart3}
\end{equation}

so that the final result, combining the results of the previous products with this last one, becomes
\begin{equation}
\frac {A_{II} (\phi_- + 1) A_{II} (\phi_- -1)} {\big( A_{II} (\phi_-) \big)^2} = (1-\phi_-)^{-2}  ( \lvert \Delta N_F^{-} \rvert + 1 -\phi_-)^{2} \cdot \bigg[ \textnormal{expression in Eq. (\ref{prodIIpart3})} \bigg]
\end{equation}

Similar expressions can be obtained for $A_{V}$ and $A_{IV}$  which follow in a similar vein to the one presented here, and will therefore be omitted. Not surprisingly, the similarity in the algebraic expressions of the various factors does lead to some quite significant simplifications, in addition to a symmetry under the combined $\Delta N_F^{\iota} \leftrightarrow \Delta N_F^{-\iota}$ and $\iota \phi_{\iota} \leftrightarrow -\iota \phi_{-\iota}$ substitutions. The resulting expression after taking all six product regions into account (I - VI), becomes 
\begin{equation}
\frac {\phi_+^2 \phi_-^2} {\big(  \lvert \Delta N_F^{+} \rvert - \phi_+ \big)^2 \big( \lvert \Delta N_F^{-} \rvert + \phi_- \big)^2} \prod_{i=\lvert \Delta N_F^{-} \rvert + 1}^{M-\lvert \Delta N_F^{+} \rvert} \left( \frac {\sin \bigg( \frac {\pi i} L- \frac {\pi} L + \frac {\pi} L  \Phi_i \bigg) \sin \bigg( \frac {\pi i} L - \frac {\pi M} L + \frac {\pi} L  \Phi_i \bigg)} {\sin \bigg( \frac {\pi i} L + \frac {\pi} L  \Phi_i \bigg) \sin \bigg( \frac {\pi i} L - \frac {\pi} L (M+1) + \frac {\pi} L  \Phi_i \bigg)} \right)^2 \label{almost}
\end{equation}

In order to have a workable expression for the recurrence relation, we need to be able to evalute the product of Eq. (\ref{almost}), which poses a problem as it contains a macroscopic number of factors. However, this is really only an issue if we were to keep the momentum dependence of the phase shift as they are written above, i.e. by strictly keeping $\Phi (q_i)$, where $q_i \sim 2\pi i / L$. In the thermodynamic limit, the pseudofermion momentum values form a continuum, and the phase shift of individual pseudofermions inside the occupied sea ceases to carry information about the system, except at the Fermi points which \emph{define} the occupied sea of pseudofermions. Therefore, the approximation $\Phi(q_{i+1}) \approx \Phi(q_i)$ is not as crude as it seems to imply, since $\Delta \bar{q}_i = 2\pi / L + (2\pi / L) (\Phi_{i+1} - \Phi_i) \approx 2 \pi / L + (2 \pi / L )^2 \Phi ' (q_i)$, meaning that between neighbouring pairs of pseudofermions, this approximation is of the order of $(1/L)$. \label{oneoverlsq} Since magnetisation $m \rightarrow 0$, we guarantee that $\Phi'(q_i)$ is a smooth function by staying away from the edges of the occupied sea, where $\Phi (q_i)$ can have sudden jumps in its value (see, for example, appendix B in Ref. \cite{carmelo62} where numerical values are given for this limit, and Figs. (\ref{phisc}) and (\ref{phiss}) where the kinks of the phase shifts are explicitly shown). Therefore, after some algebraic manipulation, the unevaluated product in Eq. (\ref{almost}) becomes
\begin{equation}
\frac {\big( \frac {\pi} L \big)^2 \big( \lvert \Delta N_F^{-} \rvert+ \phi_- \big)^2 } {\sin^2 (\pi n) + \mathcal{O}(1/L)} \cdot \frac {\big( \frac {\pi} L \big)^2 \big( \lvert \Delta N_F^{+} \rvert -  \phi_+ \big)^2 } {\sin^2 (\pi n) + \mathcal{O}(1/L)}
\end{equation}

which when putting everything together gives us a compact form of the recurrence relation of zeroth order
\begin{equation}
\frac {A( \{ \Phi + 1 \}) A( \{ \Phi - 1 \}) } {\big( A( \{ \Phi \}\big)^2} = \frac {\pi^4 \phi_+^2 \phi_-^2} {L^4 \sin^4 \pi n} \label{finalA}
\end{equation}

which is the final form of the recurrence relation we will be attempting to solve in the following section. The leading order $(1/L)$ correction term of this expression, is proportional to $\cot \pi n$, which informs us that Eq. (\ref{finalA}) is a valid expression as long as we stay away from half-filling, $n \rightarrow 1$. To conclude however, we should note that this expression readily compares to that of Ref. \cite{penc3}, where a similar recurrence relation is evaluated. In that reference, the expression obtained is similar to the one obtained here, but also fundamentally different given that in the $U/t = \infty$ limit, a constant phase shift $Q$ for all "spinless fermions" (corresponding to the $c$ pseudofermions treated here) is assumed. Furthermore, their "spinless fermion" momentum is defined such that each such quantum objects has a $Q=\pm \pi$ phase shift in its ground state, and not $0$ as in the PDT. This means that our recurrence relation gives zero if the phase shifts vanish, while the corresponding relation in that reference vanishes when $Q=\pm \pi$. Furthermore, since the spin part in the $U/t = \infty$ limit is treated as a Heisenberg spin chain, where in contrast our $s$ pseudofermions stand on equal footing to the $c$ pseudofermions, our comparisons are strictly only valid to the expression obtain for the "spinless fermions" in Ref. \cite{penc3}. Actually, the result of that reference becomes identical to the one presented here, provided that we identify $\phi_+$ with $(Q/2\pi) + 1/2$, and  $\phi_-$ with $(Q/2\pi) - 1/2$. We shall see in the next section why indeed this identification is the correct one.

\section{Closed form expression of the one electron spectral function}

\subsection{Solving the recurrence relations}
Let now $A^{(0,0)} (\phi_-,\phi_+) = \prod_{\iota=+,-} \bar{A} (\iota \phi_{\iota})$, where $ \bar{A}(\iota \phi_{\iota}) = A_0^{\iota} (\iota \phi_{\iota}) g_{\iota} (\iota \phi_{\iota})$, where the newly introduced functions $A_0$ and $g(x)$ satisfy
\begin{equation}
\begin{split}
\frac {A_0^{\iota} (\iota \phi_{\iota}+1) A_0^{\iota} (\iota \phi_{\iota}-1)} {\big( A_0^{\iota} (\iota \phi_{\iota}) \big)^2} & = \frac 1 {\big( L \sin \pi n \big)^2} \ \text{ , and}\\
& \\
\frac {g_{\iota}(\iota \phi_{\iota}+1) g_{\iota}(\iota \phi_{\iota}-1)} {\big( g_{\iota}(\iota \phi_{\iota}) \big)^2} & = \big( \pi \phi_{\iota} \big)^2.
\end{split}
\end{equation}
\newline

\begin{figure}[h!]
  \includegraphics[width=0.5\linewidth]{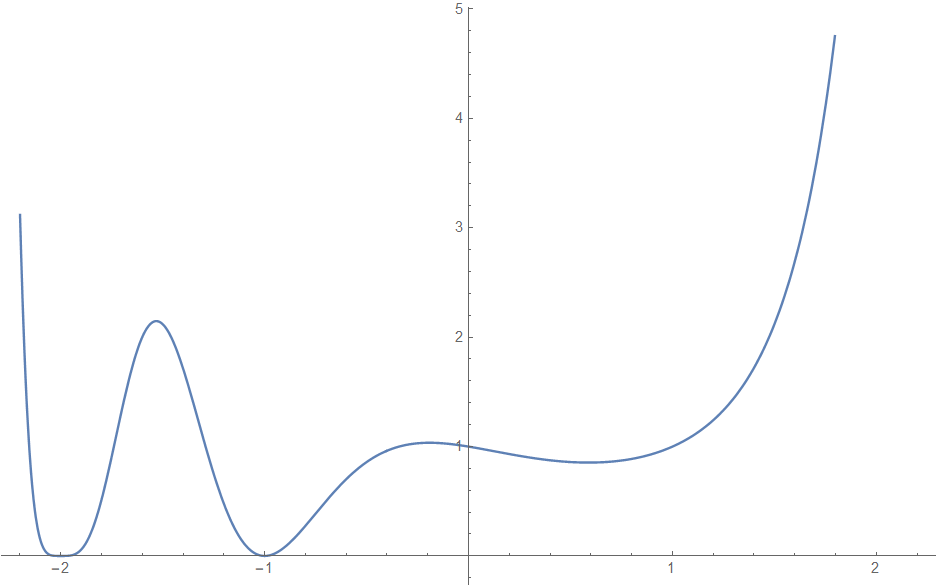}
  \centering
  \caption{The function $\pi^{x \big( x-1 \big)} \Big[ G\big( 1+x \big) \Big]^2$. It goes to infinity as $x \rightarrow \infty$ but is zero for negative integer values, with local maxima in between. For example, there is a local maximum at $x=-2.59365$, with a value of $144.512$ (not shown).}
  \label{pibarnes}
\end{figure}

Solving for $A_0 (\iota\phi_{\iota})$, we obtain
\begin{equation}
A_0^{\iota} (\iota \phi_{\iota}) = D_{1 \iota} \ D_{2 \iota}^{\iota \phi_{\iota}} \ \bigg( L \sin \pi n \bigg)^{\iota\phi_{\iota} (1-\iota\phi{\iota})} \label{Afunc}
\end{equation}
for some unknown constants $D_{1 \iota}$ and $D_{2 \iota}$ to be determined later. Next we solve the recurrence relation for the function $g(\iota\phi_{\iota})$, which can be compared to the function $f(Q)$ in Ref. \cite{penc3}: 

\begin{equation}
g_{\iota}(\iota\phi_{\iota}) = C_{1\iota} \ C_{2\iota}^{\iota\phi_{\iota}} \pi^{\iota\phi_{\iota} \big(\iota\phi_{\iota}-1 \big)} \Big[ G\big( 1+\iota\phi_{\iota} \big) \Big]^2 \label{gfunc}
\end{equation}
where $G$ is Barne's G-function. This function, with the constants omitted, is plotted in figure \ref{pibarnes}. Barne's G-function satisfies the recurrence relation $G(1+x) = \Gamma (x) G(x)$.
\newline

In order to decide the unknown constants in Eqs. (\ref{Afunc}) and (\ref{gfunc}), we will use their known values in the $U/t \rightarrow \infty$ limit (see for example Ref. \cite{penc3}). However, before the constants $C_1$ and $C_2$ can be determined, we need to understand how the phase shifts in our formalism can be compared to the phase shift $Q$ found in that reference. Indeed, in our formalism we have two phase shifts, $\phi_-$ and $\phi_+$, while Ref. \cite{penc3} only has one, $Q$. In the following, we shall see that there is a specific and consistent choice of $\phi_-$ and $\phi_+$ which reproduces all previously obtained expressions. The purpose of this check is to be able to conclude that the parameter $Q$ indeed does correspond to a physical choice of phase shifts in our framework. 
\newline

First, both frameworks need to satisfy the same boundary conditions. Indeed, in that reference's formalism we have $f(\pm \pi)=1$, where the condition $Q=\pi$ and $Q=-\pi$ is equivalent to $Q-\pi=0$ and $Q+\pi=0$. Since our phase shift are expressed in units of $2\pi$, we express these equations as $ \big( Q / 2\pi \big) - 1/2 = 0$ and $\big( Q / 2\pi \big) + 1/2 = 0$. The phase shifts in our formalism should, in the $U/t \rightarrow \infty$ limit, satisfy the same two equations. This means that by letting $\phi_- = \big( Q / 2\pi \big) - 1/2$ and $\phi_+ =  \big( Q / 2\pi \big) + 1/2$, we ensure ourselves that by choosing $Q = \pm \pi$, we obtain the same physics as by choosing $\phi_{\pm} = 0$ in our framework. For example, we notice that $\pi^4 \phi_+^2 \phi_-^2 = (\pi^2-Q^2)^2 / 2^4$ which ensures us that
\begin{equation}
\lim_{U/t \rightarrow \infty} g_{-}(-\phi_-) g_{+}(\phi_+) = g_{-}\bigg( - \frac Q {2\pi} + \frac 1 2 \bigg) \ g_{+}\bigg( \frac Q {2\pi} + \frac 1 2 \bigg) = f(Q) \label{defg}
\end{equation}
in full agreement of the expressions obtained in that reference.
\newline

Next, we will use some known values of the function $f(Q)$ to determine some of the unknown constants of the function $g$. First, $f(\pm \pi)=1$ translates in our model to having $Q=\pi$ when used in $\phi_- $ (hence giving $\phi_-  = 0$), and $Q=-\pi$ when used in $\phi_+$ (giving $\phi_+ = 0$):
\begin{equation}
1 = g_{+}(0) g_{-} (0)  \Rightarrow  C_{1+} C_{1-} = 1
\end{equation}

Next, $f(0) \approx 0.737391$. In our formalism, we see that 
\begin{equation}
\begin{split}
f(0) & = g_{-}(1/2) \cdot g_{+} (1/2) = C_{1+} C_{1-} \big( C_{2-} C_{2+} \big)^{1/2} \pi^{-1/2} G^4 (3/2) =  \\
& = \frac {2^{1/6} \sqrt{\text{e} \pi}} {A^6} \cdot \big( C_{2-} C_{2+} \big)^{1/2}  \approx 0.737391 \cdot \big( C_{2-} C_{2+} \big)^{1/2} 
\end{split}
\end{equation}
where $A \approx 1.28243$ is Glaischer's constant. Interestingly, Ref. \cite{penc3} produces the number $0.737391$ as an approximation to $f(0)$, but due to the known exact values of Barne's function we can conclude that we must have $ C_{2-} C_{2+}=1$ for the two functions to be consistent with each other. Note that even though $\phi_- = -1/2$ when $Q=0$, the argument of $g_-$ is by definition the negative of this value, $+1/2$. We thus conclude that $g(x) = \pi^{x (x-1)} G(1+x)^2$.
\newline

Putting the results together, we obtain,
\begin{equation}
A^{(0,0)} (\phi_-,\phi_+) = \prod_{\iota=+,-}  \bar{A} (\iota \phi_{\iota}) = \prod_{\iota=+,-} \alpha_{1 \iota} \  \alpha_{2 \iota}^{\iota \phi_{\iota}} \ \bigg( L \sin \pi n \bigg)^{\iota\phi_{\iota} (1-\iota\phi{\iota})} \pi^{\iota\phi_{\iota} \big(\iota\phi_{\iota}-1 \big)} \Big[ G\big( 1+\iota\phi_{\iota} \big) \Big]^2 \label{Azerozero}
\end{equation}
for some new constants $\alpha_{1 \iota}$ and $\alpha_{2 \iota}$. One thing to notice is that the combined exponent of the factor $L \sin \pi n$, namely $-\phi_- (1+\phi_-) + \phi_+ (1-\phi_+)$ is exactly equal to the exponent in the corresponding expression Eqs. (24) and (25) of Ref. \cite{penc3}. This can be readily be verified if we substitute the expressions of $\phi_-$ and $\phi_+$ in terms of $Q$ into the expression of the exponent. We thus have a complete agreement with the $U/t \rightarrow \infty$ results.
\newline

\subsection{Particle-hole processes}
The lowest peak weight obtained in the previous subsection (up to some unknown constants), $A^{(0,0)}$ of Eq. (\ref{Azerozero}), gives the spectral weight associated with the ground state $\rightarrow$ final state transition upon removing or adding one electron. However, there exists a set of eigenstates with energy and momentum in the $(1/L)$ vicinity of the point in the $(k,\omega)$ plane where $A^{(0,0)}$ have allocated its spectral weight, namely the particle-hole processes in the low-lying linear regime. Indeed, the superscript $(0,0)$ stands for "zero height" at the left and right Fermi points respectively, equal to the number of steps (in units of $2 \pi /  L$) that particle-hole processes would bring the momentum and energy of the excited eigenstate to. Following this notation, $A^{(i,j)}$ is the spectral weight allocated at a neighbouring point in the plane, with momentum $k+(2\pi/L)(j-i)$ and energy $\omega + (2\pi v /L) (i+j)$, where $v$ is shorthand for $v_{\alpha} (q_{F\alpha})$, the group velocity at the Fermi level. The total momentum of the particle-hole processes on the left (-) Fermi point is thus $(-2\pi i / L)$, and correspondingly the total momentum of the particle-hole processes on the right (+) Fermi point is $(+2\pi j / L)$. 
\newline

We let the particle-hole processes around the left and right Fermi points have total momentum and energy given by $(k_{\alpha}, \omega_{\alpha})$ for each pseudofermion branch, $\alpha = c, s$. In the following, we outline the main points in the derivation of the spectral function (see, for example, Ref. \cite{carmelo64} for a more thorough treatment). This is justified considering the simplified regime relevant to the studies in this paper, since we are only considering excited final states involving occupancies of $c$ and $s$ pseudofermions, where $m \rightarrow 0$. 
\newline

The overall one electron spectral function can now be written as (dropping the superscript $\pm$):
\begin{equation}
A(k,\omega) = \sum_{\text{f}} \sum_{\text{ph}} \delta_{k , P(q_c,q_s) + \Delta k} \, \delta_{\Delta k , k_c +k_s} A_c \big( k_c, \omega_c \big) A_s \big( k_s,\omega_s \big) \delta (\omega - E(q_c,q_s) - \Delta \omega) \delta (\Delta \omega - \omega_c - \omega_s) \label{spectral}
\end{equation}

where $\Delta \omega=\omega_c + \omega_s$ and $\Delta k = k_c + k_s$ is the energy and momentum of the particle-hole excitations, while $E(q_c,q_s)$ and $P(q_c,q_s)$ are the finite energy and momentum values at the base of the particle-hole towers, given by Eq. (\ref{setup}) by replacing $k$ and $\omega$ in that equation with $E(q_c,q_s)$ and $P(q_c,q_s)$, respectively. A typical final state brings the excitation energy to $E(q_c,q_s)$, upon which energy from the linear particle-hole regime $\Delta \omega$ is added, allocating spectral weight at $\omega = E(q_c,q_s) + \Delta \omega$. For example, in the one-electron addition case, for which one $c$ pseudofermion and one $s$ pseudofermion-hole is created, we have $E(q_c,q_s) = \varepsilon_{c}^0 (q_{c}) - \varepsilon_{s}^0 (q_{s})$ and $P(q_c,q_s) =q_{c}-q_{s}$, respectively. The particle-hole (ph) summation accounts for all configurations of occupancies/unoccupancies of particle-hole character within the linear regime of both the left and right Fermi points. Simplifications can be made in the $U/t \rightarrow \infty$ limit where the $s$ pseudofermion dispersion is flat, rendering $v_s (q_{Fs}) = 0$. In addition, the zero magnetisation limit ensures that the entire $s$ band is filled, leaving only a vanishing space left for unoccupied momenta values above the Fermi level being reachable by particle-hole excitations in that case. However, we will aim to keep the discussion general with a finite group velocity and with a magnetisation which is not strictly equal to zero.
\newline

Important to note is that a pseudofermion particle-hole excitation within the low-lying regime does not change its phase shift, as all phase shifts can be approximated by $\Phi_{\alpha} (q_{F\alpha})$ in that regime, with a correction term of order $\mathcal{O}(L^{-2})$, as concluded by following a similar reasoning as on p. \pageref{oneoverlsq}. However, the particles created due to particle-hole processes still scatter with the finite-energy and momentum excitations made by the operator $\hat{\Theta}(k)$ in Eq. (\ref{slater2}), which as we shall see leads to a power-law decay of the spectral weight as we move away from singular features on the $(k,\omega)$ plane, manifesting the underlying Anderson orthogonal catastrophe. This means that the $\alpha$-factorised part of the spectral function, $A_{\alpha} (k_{\alpha},\omega_{\alpha})$, can be written as an overlap between two densely packed states as before, but with a decaying part multiplying the lowest peak weight corresponding to a tower of particle-hole states in the low-lying linear regime.

\subsection{Particle-hole processes: relative weights}
Since the final state occupancies of pseudofermions are so similar between (1) a typical lowest peak final state and (2) an eigenstate with a particle-hole excitation existing on top of this lowest peak weight, it is possible to exactly evalute the contribution from the particle-hole excitation in question in terms of said lowest peak weight. In fact, since the expression for the lowest peak weight, Eq. (\ref{slater2}), consists of a string of factors where each factor corresponds to an individual momentum position, we can easily swap one such occupied momentum value (within the linear regime of either the left or right Fermi point, thus creating a hole (h) at that position), with another unoccupied momentum value also within the linear regime (creating a particle (p) at that momentum value), and then compare it with the corresponding expression without such a swap. Before we continue the analysis, a brief comment of the nature of the ground state is in order. When evaluating the matrix overlap that occurs in the expression for the spectral function, our ground state was already modified by a finite number of pseudofermions being created or annihilated at the Fermi points defining the seas of occupied momentua values for each pseudofermion. This choice was done due to computational convenience when evaluating the Slater determinant of Eq. (\ref{slater}), since in order to evaluate the total spectral function, such zero-energy processes must be taken into account anyhow. However, the definition of the relative weights considered here is such that it compares the final excited eigenstate (with a finite number of particle-hole pairs) with the unshifted "true" $N_{\alpha}$-particle ground state of the system. In other words, by letting $\hat{\Theta}_F$ be the operator acting onto the original $N_{\alpha}$-particle ground state generating the zero-energy processes, we have $\lvert \widetilde{GS} \rangle = \hat{\Theta}_F \lvert GS \rangle$. As before, we may also choose to have this operator act on the left (the bra) instead of the right (the ket), equivalent of generating the conjugate of such zero-energy processes on the excited final state. Since the occupied momenta values of the final state have twisted boundary conditions as shown by the non-zero phase shift values, this process will yield finite-size corrections compared to the usual $(2\pi /L)$ momentum spacing involving both the number of created and annihilated pseudofermions at the Fermi points, $\Delta N_{F\alpha}^{\pm}$, and the corresponding phase shifts at the Fermi point.
\newline

Assume $N_{ph,\iota}$ is the number of particle-hole pairs created at the $\iota=\pm$ Fermi point. This leads to an expression of the spectral weight of the form $A(N_{ph}) = a_{N_{ph,+}}^+ a_{N_{ph,-}}^- A^{(0,0)}$, where $A^{(0,0)}$ is given by Eq. (\ref{slater2}) and where $a_{N_{ph,\iota}}^{\iota}$ is the \emph{relative weight} corresponding to particle-hole excitations near the $\iota=\pm$ Fermi point. Some expressions of this relative weight include, $a_0^{\pm} = 1$ (trivially), and $a_1^{\pm} = \gamma (p_1,h_1)$, where
\begin{equation}
\gamma(p_1,h_1) = \prod_{\substack{i=1 \\ i \neq h1}}^M \frac {\sin^2 \Big( \frac 1 2 \big[ \bar{q}_i - \bar{q}_{p1} \big] \Big)} {\sin^2 \Big(\frac 1 2 \big[ \bar{q}_i - \bar{q}_{h1} \big] \Big)} \prod_{i=1}^N \frac {\sin^2 \Big( \frac 1 2 \big[ q_i - \bar{q}_{h1} \big] \Big)} {\sin^2 \Big( \frac 1 2 \big[ q_i - \bar{q}_{p1} \big] \Big)}.
\end{equation}

where the barred momenta values correspond to pseudofermions of the final excited state, and the un-barred momenta values to the initial ground state (with no phase shift). For example, $\bar{q}_{p1}$ is the momentum value of the created pseudofermion, originating from a momentum value $\bar{q}_{h1}$ where there is now a pseudofermion hole. These indeces are chosen such that $h$ corresponds to pseudofermion hole momentum value $q_{F\alpha}^{\pm} \mp (2\pi /L)(h-1)$, and $p$ corresponds to pseudofermion particle momentum value $q_{F\alpha}^{\pm} \pm (2\pi /L)p$, respectively. 

The relative weight for the case of two particle-hole excitations is given by:
\begin{equation}
a_2^{\pm} = \frac {\sin^2 \Big( \frac 1 2 \big[ \bar{q}_{p1} - \bar{q}_{p2} \big] \Big) \sin^2 \Big(\frac 1 2 \big[ \bar{q}_{h1} - \bar{q}_{h2}\big] \Big)} {\sin^2 \Big( \frac 1 2 \big[ \bar{q}_{p1} - \bar{q}_{h2} \big] \Big) \sin^2 \Big(\frac 1 2 \big[ \bar{q}_{h1} - \bar{q}_{p2}\big] \Big)} \cdot \gamma(p_1,h_1) \gamma(p_2,h_2) \label{a2}
\end{equation}
where $h1$ and $h2$ correspond to the indeces of the two holes, and correspondingly, $p1$ and $p2$ for the two particles (both of which should be in the vicinity of the same $\pm$ Fermi point). This pattern continues for eigenstates with more particle-hole pairs, where it can be shown that in the general case we have:
\begin{equation}
a_{N_{ph}}^{\pm} = \prod_{\mathcal{C}_{ij}} \frac {\sin^2 \Big( \frac 1 2 \big[ \bar{q}_{pi} - \bar{q}_{pj} \big] \Big) \sin^2 \Big(\frac 1 2 \big[ \bar{q}_{hi} - \bar{q}_{hj}\big] \Big)} {\sin^2 \Big( \frac 1 2 \big[ \bar{q}_{pi} - \bar{q}_{hj} \big] \Big)} \Bigg|_{i \ne j} \cdot \prod_{i=1}^{N_{ph}} \gamma(p_i,h_i) \label{relnph}
\end{equation}
where $\mathcal{C}_{ij}$ are all combinations of the indices $i$ and $j$ within each sine-function (note however, that the product disregards factors for which $i=j$).
\newline

The factors $\gamma(p_i,h_i)$ simplify in the thermodynamic limit, where according to the same limiting procedure as described on p. \pageref{limitproc}, we obtain
\begin{equation}
\begin{split}
\gamma(p_i,h_i) \rightarrow & \frac {\Psi_{\iota}^2} {(p_i+h_i-1)^2} \cdot \prod_{\substack{j=1 \\ j \neq h_i}}^{p_i+h_i-1} \frac {(j-h_i+\iota \Psi_{\iota})^2} {(j-h_i)^2}, \, \, \, L \rightarrow \infty. \\
\Psi_{\iota} &= \Delta N_{F\alpha}^{\iota} + \iota \phi_{\iota} \\
\end{split} \label{relwght}
\end{equation}
Since there are in general many combinations of particle-hole pairs giving the same momentum and energy within the linear regime, this expression hides a degeneracy which grows rapidly with $N_{ph,\iota}$. Only the very lowest particle-hole excitations are easy to write down explicitly, however it is instructional to do so even though we will "not need" to write any of these down as we shall see shortly. 
\newline

Consider one particle-hole pair at the right Fermi point, originating \emph{at} the Fermi point itself, being excited to just one $(2 \pi / L)$ step above the Fermi level. This would yield a relative weight $a_1^+ = \Psi_+^2$ in the thermodynamic limit, as per Eq. (\ref{relwght}). Similarily, at the left Fermi point we would obtain the relative weight $a_1^- = \Psi_-^2$ according to the same equation. Next, consider a combined particle-hole excitation where the previous two excitations happen simultaneously: the relative weights combine to form $\Psi_+^2 \Psi_-^2$, in other words the total spectral weight would be $A^{(1,1)} =\Psi_+^2 \Psi_-^2 A^{(0,0)}$. This combined particle-hole excitation has energy $2\cdot (2\pi v / L)$ and momentum $(2\pi / L) - (2\pi / L) = 0$. 
\newline

Consider another example where a particle-hole excitation at the right Fermi point has energy $2\cdot (2\pi v / L)$ and momentum $2 \cdot (2\pi / L)$, corresponding to one pseudofermion being raised two $(2\pi / L)$ steps. There are two possibilities: A pseudofermion originating at the Fermi point with momentum $q_{F\alpha}$ attaining energy $2\cdot (2\pi v / L)$ and momentum $q_{F\alpha} + 2 \cdot (2\pi / L)$, or a pseudofermion originating at $q_{F\alpha} - (2\pi / L)$ being raised two steps, increasing its energy and momentum by the same amounts, but being $(2\pi v / L)$ and $q_{F\alpha} + (2\pi / L)$, respectively. The same situation would occur at the left Fermi point, where the momentum values decrease by the same values. This leads to a total relative weight at the $\iota$ Fermi point
\begin{equation}
a_1^{\iota} = \frac {\Psi_{\iota}^2} {2^2} \frac {(1+\Psi_{\iota})^2} {1^2} + \frac {\Psi_{\iota}^2} {2^2} \frac {(-1+\Psi_{\iota})^2} {1^2} = \frac {\Psi_{\iota}^2} {2} (1+ \Psi_{\iota})^2 \label{a1}
\end{equation}
meaning that $A^{(0,2)} =\big[ \Psi_{+}^2 (1+\Psi_{+})^2 / 2 \big] A^{(0,0)}$, and similarily for $A^{(2,0)}$ (by replacing $\Psi_+$ with $\Psi_-$). A similar situation occurs for $A^{(0,3)}$ and $A^{(3,0)}$ involving three possible particle-hole excitations, leading to the final expression $A^{(0,3)} =\big[ \Psi_{+}^2 (1+\Psi_{+})^2 (2+\Psi_{+})^2 / 2\cdot 3 \big] A^{(0,0)}$, and similarily for $A^{(3,0)}$ (by replacing $\Psi_+$ with $\Psi_-$). 
\newline

Composite left/right particle-hole excitations can be built in a similar way by combining the possible particle-hole configurations that lead to a desired total energy and momentum. For example, $A^{(1,2)}$ involve the $A^{(0,2)}$ contribution from Eq. (\ref{a1}) combined with $a_1^- = \Psi_-^2$, giving us $A^{(1,2)} = \big[ \Psi_-^2 \Psi_{+}^2 (1+\Psi_{+})^2 / 2 \big] A^{(0,0)}$, while conversely $A^{(2,1)} = \big[ \Psi_+^2 \Psi_{-}^2 (1+\Psi_{-})^2 / 2 \big] A^{(0,0)}$.
\newline

Before we move on, one final example with $N_{ph,\iota} = 2$ is in order. The lowest possible particle-hole excitation involving two particle-holes at the same Fermi point corresponds to the spectral weights $A^{(0,4)}$ and $A^{(4,0)}$. As with $A^{(0,2)}$ and $A^{(0,3)}$ (or correspondingly $A^{(2,0)}$ and $A^{(3,0)}$ at the left Fermi point), we can raise one single pseudofermion 4 steps (there are 4 possible such $N_{ph,\iota} = 1$ particle-hole excitations), but we can also have two pseudofermions jumping two $(2\pi / L)$ steps each, above the right or left Fermi points, respectively. This means that the pre-factor of Eq. (\ref{a2}) comes into play, yielding
\begin{equation}
\frac {\sin^2 (\frac {\pi} L) \sin^2 (\frac {\pi} L)} {\sin^2 (\frac {2\pi} L) \sin^2 (\frac {2\pi} L)} \rightarrow \frac 1 {2^2} \cdot \frac 1 {2^2} = \frac 1 {16}, \, \, \, L \rightarrow \infty. \label{nph2}
\end{equation}

and we thus obtain (where we choose $\phi_+$ for $a_2^+$ and $-\phi_-$ for $a_2^-$):
\begin{equation}
a_2^{\pm} = \frac 1 {16} \gamma(1,1)\gamma(2,2) + \gamma(4,1) + \gamma(3,2) + \gamma(2,3) + \gamma(1,4) = \frac {\Psi_{\iota}^2 (1+\Psi_{\iota}^2) (2+\Psi_{\iota}^2) (3+\Psi_{\iota}^2)} {1 \cdot 2 \cdot 3 \cdot 4}.
\end{equation}

This procedure can be generalised to any spectral weight within the linear regime, where for example a combined left/right particle-hole excitation having momentum $(-2\pi i / L)$ on the left side, and $(+2\pi i' / L)$ on the right, leads to a spectral weight $A^{(i,i')} = a_{N_{ph,+}}^+ a_{N_{ph,-}} A^{(0,0)}$, where the combined relative weight is
\begin{equation}
a_-(i) \cdot a_+(i') = \frac {A^{(i,i')}} {A^{(0,0)}} = \frac {\Psi_{-}^2 (1+\Psi_{-}^2) (2+\Psi_{-}^2) \ldots (i-1+\Psi_{-}^2)} {i !} \cdot \frac {\Psi_{+}^2 (1+\Psi_{+}^2) (2+\Psi_{+}^2) \ldots (i'-1+\Psi_{+}^2)} {i' !}. \label{aii1}
\end{equation}
defining the new quantity $a_{\iota}(i)$.
\newline

We notice now that the expression in Eq. (\ref{aii1}) can be summarised by using $\Gamma$ functions, which in turn approximate to power-law expressions as $i$ and $i'$ grow beyond the first few particle-hole positions:
\begin{equation}
a_-(i) a_+(i') = \frac {\Gamma(i+\Psi_-^2)} {\Gamma(i+1) \Gamma(\Psi_-^2)} \cdot \frac {\Gamma(i'+\Psi_+^2)} {\Gamma(i'+1) \Gamma(\Psi_+^2)} \approx \frac 1 {\Gamma(\Psi_-^2) \Gamma(\Psi_+^2)} \Big( i + \frac {\Psi_-^2} 2 \Big)^{\Psi_-^2 - 1} \Big( i' + \frac {\Psi_+^2} 2 \Big)^{\Psi_+^2 - 1}
\end{equation}

This expression is identical to the previously obtained $U/t \rightarrow \infty$ result of Refs. \cite{penc1}-\cite{penc4}, using the substitution $\Psi_- = (Q/2\pi) - 1/2$ and  $\Psi_+ = (Q/2\pi) + 1/2$, for $\Delta N_{F\alpha}^{\pm} = 0$ (which is justified since the authors of those references studied a finite-sized system where no separate zero-energy processes occurred). We notice as expected that the approximation is not valid as $\Psi_{\iota} \rightarrow 0$, or in other words as the exponents in the expression above approaches $-1$, as the non-interacting case should only produce $\delta$-function peaks in the spectral weight.

\subsection{Final closed form expression}
Summarising our results so far, the pseudofermion spectral function $A_{\alpha} (k_{\alpha},\omega_{\alpha})$ is associated with a matrix overlap between the $\alpha$ pseudofermion ground state and an excited energy eigenstate involving pseudofermion particle-hole excitations on the left $(\iota= -)$ and the right $(\iota=+)$ Fermi points, respectively. The base of this tower of states is denoted by $A^{(0,0)}_{\alpha}$, and we obtain the $A_{\alpha} (k_{\alpha},\omega_{\alpha})$ spectral function by summing over all particle-hole processes with energy $(2\pi v_{\alpha} /L)(i+i')$ and momentum $(2\pi / L) (i'-i)$, respectively:
\begin{equation}
A_{\alpha}(k_{\alpha},\omega_{\alpha}) = \sum_{i,i'} A^{(0,0)}_{\alpha} a_{+}(i') a_{-}(i) \delta \left( \omega_{\alpha} - \frac {2\pi v_{\alpha}} L (i+i') \right) \frac {2 \pi} L \delta \left( k_{\alpha} - \frac {2\pi} L (i'-i) \right). \label{phstates}
\end{equation}

where the discrete-to-continuous limit contributes through the momentum $\delta$ functions according to $\delta_{k,k'} \rightarrow (2\pi / L) \delta (k-k')$. By employing the $\delta$ functions in the above summations, we obtain
\begin{equation}
A_{\alpha}(k_{\alpha},\omega_{\alpha}) = A^{(0,0)}_{\alpha} \frac {L} {8 \pi v_{\alpha}}  \prod_{\iota=\pm} a_{\iota} \left( \frac {\omega_{\alpha} + \iota v_{\alpha} k_{\alpha}} {4 \pi v_{\alpha} / L} \right) = A^{(0,0)}_{\alpha} \frac {L} {8 \pi v_{\alpha}}  \prod_{\iota=\pm} \left( \frac {\omega_{\alpha} + \iota v_{\alpha} k_{\alpha}} {4 \pi v_{\alpha} / L} + \frac {\Psi_{\iota}^2} 2\right)^{\Psi_{\iota}^2 - 1} \Big/ \Gamma(\Psi_{\iota}^2) \label{fullsp}
\end{equation}
where $A^{(0,0)}_{\alpha}$ is given by Eq. (\ref{Azerozero}). Now, in the thermodynamic limit, we wish our final expression to be system-size independent, leading us to identify factors of $L$, according to
\begin{equation}
\prod_{\iota=+,-} \alpha_{1 \iota} \  \alpha_{2 \iota}^{\iota \phi_{\iota}} \ \bigg( L \sin \pi n \bigg)^{\iota\phi_{\iota} (1-\iota\phi{\iota})} L^{\Psi_{\iota}^2 - 1/2} = \left\{
                \begin{alignedat}{4}
                    &\alpha_{1\iota} = (L \sin \pi n )^{-(\Delta N_{F\alpha}^{\iota})^2 + 1/2} \\
                    &\alpha_{2\iota} = (L \sin \pi n)^{-(1+2\Delta N_{F\alpha})} \\
                \end{alignedat}
                 \right\} = \big( \sin \pi n \big)^{-\Psi_{+}^2 - \Psi_{-}^2+1/2}
\end{equation}
\newline

which leads to
\begin{equation}
\begin{split}
A_{\alpha}(k_{\alpha},\omega_{\alpha}) &= \frac 1 {4\pi v_{\alpha} \sin \pi n_{\alpha}} \prod_{\iota = \pm} \frac {g(\iota \phi_{\iota})} {\Gamma(\Psi_{\iota}^2)} \left( \frac {\omega_{\alpha} + \iota v_{\alpha} k_{\alpha}} {8 \pi v_{\alpha}\sin \pi n_{\alpha}} \right)^{\Psi_{\iota}^2 - 1} \\
\\
g(x) &= \pi^{x \big(x-1 \big) } \bigg[ G(1+x)\bigg]^2 \\
\\
\phi_{\iota} &= \Phi_{\alpha} (q_{F\alpha}^{\iota}) = \phi_0 + \Phi_{\alpha, c} (q_{F\alpha}^{\iota},q_{c}) \Delta N_{c} (q_{c}) + \Phi_{\alpha, s} (q_{F\alpha}^{\iota},q_s) \Delta N_{s} (q_{s}) \\ 
\\
\Psi_{\iota} &= \Delta N_{F\alpha}^{\iota} + \iota \phi_{\iota} \\
\label{final}
\end{split}
\end{equation}

for the expression of the one electron spectral function (where $n_{\alpha}$ is the filling of the $\alpha$ pseudofermion, and $\phi_0 = -1/2,0,1/2$), which is the general finite $U/t$ result equivalent to Eq. (33) of Ref. \cite{penc3}, which states the corresponding equivalent expression for the limit $U/t \rightarrow \infty$.
\newline

\subsection{Numerical results and concluding comments}
\begin{figure}[h!]
  \includegraphics[width=0.99\linewidth]{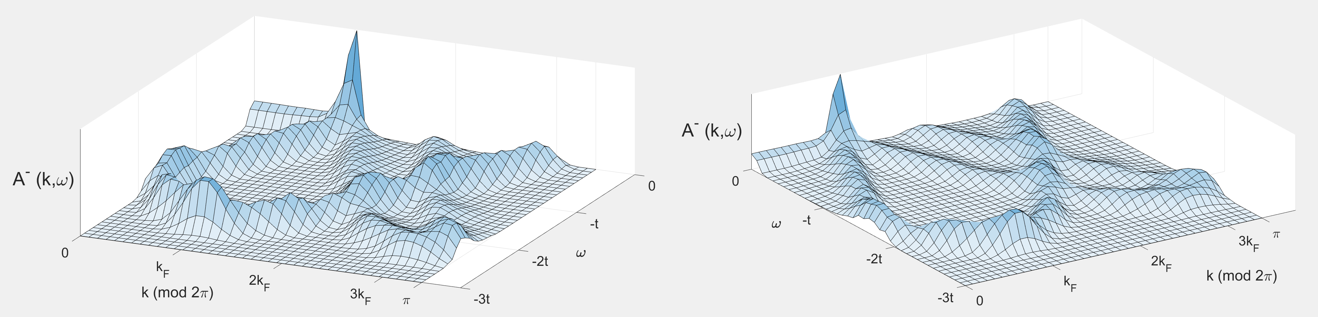}
  \centering
  \caption{The one electron removal spectral function as viewed from two different angles, for $U/t = 400$, filling $n=0.59$ and magnetisation $m \rightarrow 0$, as plotted from Eq. (\ref{final}). The momentum $k$ is plotted$\! \mod 2\pi$. Small irregularities of the plotted surface are due to a small sampling space (41 x 41 points), and restrictive signal processing (a.k.a. smoothing). Notable features of the spectral map is the strong weight corresponding to either a $c$ or an $s$ pseudofermion being created at its corresponding Fermi points (especially showing a high peak at momentum $k_F = \pi n / 2$). Moreover, prominent singular features are shown as one pseudofermion is annihilated at its Fermi point while the other is annihilated along its dispersion band. We see two examples of this: one giving rise to a zero-energy ridge (s-band) between $k=0$ and $k=k_F$, corresponding to a an $s$ pseudofermion being annihilated along its nearly-identically zero energy dispersion band, and one following an arch along the $c$ dispersion relation from $k=k_F$ to $k=3k_F$. A strong van-Hove singularity is also visible at the bottom of the latter arch, in a region where $v_c (q_c) \approx 0$.} \label{Apic}
\end{figure}

Previously, a closed form expression for the function $g(x)$ of the one electron spectral function had not been found (see for example Ref. \cite{carmelo64}, Appendix J, for approximate expressions involving estimated quantities whose exact expressions were unknown). However, the critical exponents governing the power-law decay are the same in this paper as what has been previously reported \cite{carmelo61}-\cite{carmelo64}. However, with the results obtained here, we now know how to compare the $U/t \rightarrow \infty$ constant phase shift of Refs. \cite{penc1}-\cite{penc4}, with the state dependent phase shifts emerging within the PDT. Moreover, the overall shape of the spectral weight of the entire $(k,\omega)$ plane obtained here manages to reproduce previously known results, as shown in Fig. (\ref{Apic}), confirming the accuracy of our analysis. It should be stated that the results reported in this paper are only valid in the thermodynamic limit, however even though we have exemplified the evaluation of our spectral function in the $U/t \rightarrow \infty$ limit, the obtained result of Eq. (\ref{final}) is valid for all values of  $U/t > 0$. 
\newline

\begin{figure}[h!]
  \includegraphics[width=0.5\linewidth]{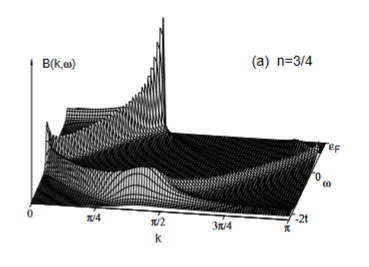}
  \centering
  \caption{The one electron removal spectral function from Ref. \cite{penc4}, showing similar features as compared to the spectral function obtained from Eq. (\ref{final}), with small differences due to a change in filling $n=0.75$ (in Fig. (\ref{Apic}), the filling is $n=0.59$). The sampling space in this picture is noticeably larger (201 x 201 points) than in Fig. (\ref{Apic}). Most notably, the singular features in this figure are somewhat "dragged out" due to having a larger Fermi momentum at $\pi n$, for example notable at the position of the strong Fermi peak at $(k,\omega)=(k_F,0)$. Furthermore, the wide arch ($c$ band) following a curve of singular features extends further for large $k$ values in this figure, which is also due to a larger filling. Hence, we see a "bounce back" along the Brillouin zone at $k=\pi$ since the dispersion band surpasses the limit of the first Brillouin zone.} \label{Karlopic}
\end{figure}

In Fig. (\ref{Apic}), the one electron removal spectral function is plotted, following Eq. (\ref{final}). The usual singular features stemming from Anderson's orthogonality catastrophe and the finite-size corrections due to the values of $\Psi_{\iota}$ are easily identified. For example, the line-shapes originating from $k=k_F$ (which is nothing but the $s$ pseudofermion Fermi point), are dominant along both pseudofermion dispersion bands. Note that due to the state-dependent phase shifts introduced within the framework of the PDT, these line-shapes give rise to a power-law decay of the spectral weight due to particle-hole excitations, as we move away from the region defined by the disperion relations. The momentum $k$ is given by $q_c+q_s-\pi n \phi_0 \, \text{mod} 2\pi$, where $q_c \in [-2k_F,2k_F]$ and $q_s \in [ -k_F, k_F ]$ and since we are plotting the spectral function in the positive half-Brilloiun zone we notice that the $c$ dispersion line shape seem to "bounce off" the lines given by $k=0$ and $k=k_F$, respectively. Equally important, and in agreement with previous results, we see a strong suppression of the spectral weight in other parts of the $(k,\omega)$ plane. See for example the line-shape following the $c$ pseudofermion dispersion starting at $k=k_F$, wrapping around $k=\pi$ to end up at $k=2k_F$. In comparison with Fig. (\ref{Karlopic}), it is worth noting that since that reference uses a slighty higher value for the electron filling, the corresponding $c$ energy band is fuller, resulting in a wider breadth in the line-shapes in the $(k,\omega)$ plane in that figure. Apart from this small discrepancy, there is a wide agreement in the overall allocation of spectral weight between the two results.

\clearpage

\begin{thebibliography}{999}

\bibitem{hubbard1}
	J. Hubbard, Royal Soc. of London A, 276(1365), 238-257 (1963)

\bibitem{mott0}
	N. F. Mott (1968) Rev. Mod. Phys. 40 677

\bibitem{mott1}
	E. Dagotto, Rev. Mod. Phys., 66(3), 763-840 (1994)

\bibitem{mott2}
	C. A. Stafford, A. J. Millis (1993) Phys. Rev. B 48 1409

\bibitem{mott3}
	H. J. Schulz (1990) Phys. Rev. Lett. 64 2831

\bibitem{mott4}
	R. J{\"o}rdens, N. Strohmaier, K. G{\"u}nter, H. Moritz, T. Esslinger (2008) Nature 455 204–7

\bibitem{mott5}
	J. Kokalj, R. H. McKenzie (2013) Phys. Rev. Lett. 110 206402

\bibitem{scsep1}
	R. Senaratne, D. Cavazos-Cavazos, S. Wang, F. He, Y-T Chang, A. Kafle, H. Pu, X-W Guan, R. G. Hulet (2022) Science 376 1305–8

\bibitem{scsep2}
	Auslaender O M, Steinberg H, Yacoby A, Tserkovnyak Y, Halperin B I, Baldwin K W, Pfeiffer L N and West K W 2005 Science 308 88–92

\bibitem{scsep3}
	Kim C, Matsuura A Y, Shen Z-X, Motoyama N, Eisaki H, Uchida S, Tohyama T and Maekawa S 1996 Phys. Rev. Lett. 77 4054

\bibitem{scsep4}
	Kim B Jet al 2006 Nat. Phys. 2 397–40

\bibitem{scsep5}
	Jompol Y, Ford C J B, Griffiths J P, Farrer I, Jones G A C, Anderson D, Ritchie D A, Silk T W and Schofield A J 2009 Science 325 597–601

\bibitem{scsep6}
	Voit J 1993 J. Phys.: Condens. Matter 5 8305

\bibitem{scsep7}
	Schmidt T L, Imambekov A and Glazman L I 2010 Phys. Rev. B 82 245104

\bibitem{scsep8}
	Lorenz T, Hofmann M, Gr¨uninger M, Freimuth A, Uhrig G S, Dumm M and Dressel M 2002 Nature 418 614–7

\bibitem{scsep9}
	Scopa S, Calabrese P and Piroli L 2022 Phys. Rev. B 106 134314

\bibitem{scsep10}
	Senaratne R, Cavazos-Cavazos D, Wang S, He F, Chang Y-T, Kafle A, Pu H, Guan X-W and Hulet R G 2022 Science 376 1305–8

\bibitem{book1}
	T. Giamarchi, "Quantum Physics in One Dimension," Oxford University Press (2004)

\bibitem{supercond1}
	M. Imada, A. Fujimori, Y. Tokura, Rev. Mod. Phys., 70(4), 1039-1263 (1998)

\bibitem{supercond2}
	Miyake K 2007 J. Phys.: Condens. Matter 19 125201

\bibitem{supercond3}
	Pruschke T, Jarrell M and Freericks J K 1995 Adv. Phys. 44 187–210

\bibitem{supercond4}
	Dagotto E 2005 Science 309 257–62

\bibitem{supercond5}
	Yanase Y, Jujo T, Nomura T, Ikeda H, Hotta T and Yamada K 2003 Phys. Rep. 387 1–149

\bibitem{TLL1}
	Luttinger J M 1963 J. Math. Phys. 4 1154–62

\bibitem{TLL2}
	C. Kane, M. P. A. Fisher, PRL 68(8), 1220-1223 (1992)

\bibitem{TLL3}
	Haldane F D M 1981 J. Phys. C: Solid State Phys. 14 2585

\bibitem{TLL4}
	Scopa S, Calabrese P and Piroli L 2022 Phys. Rev. B 106 134314

\bibitem{ultracold1}
	Gall M, Wurz N, Samland J, Chan C F and K¨ohl M 2021 Nature 589 40–3

\bibitem{ultracold2}
	I. Bloch, J. Dalibard, S. Nascimbène, Nat. Phys., 8(4), 267-276 (2012)

\bibitem{ultracold3}
	Scherg S, Kohlert T, Sala P, Pollmann F, Madhusudhana B H, Bloch I and Aidelsburger M 2021 Nat. Commun. 12 4490

\bibitem{ultracold4}
	Vijayan J, Sompet P, Salomon G, Koepsell J, Hirthe S, Bohrdt A, Grusdt F, Bloch I and Gross C 2020 Science 367 186–9

\bibitem{ultracold5}
	Koepsell J, Vijayan J, Sompet P, Grusdt F, Hilker T A, Demler E, Salomon G, Bloch I and Gross C 2019 Nature 572 358–62

\bibitem{ultracold6}
	M. Greiner et al., Nature, 415(6867), 39-44 (2002)

\bibitem{LiebWu}
	E.H. Lieb, F.Y. Wu, Phys. Rev. Lett. 20 1445 (1968)

\bibitem{takahashi}
	M. Takahashi, Prog. Theor. Phys. 47 69 (1972)

\bibitem{josenuno}
	J.M.P. Carmelo, N.M.R. Peres, Phys. Rev. B 56 3717 (1997)

\bibitem{haldane}
	F.D.M. Haldane, Phys. Rev. Lett. 67 937 (1991)

\bibitem{carmelo1}
	J. M. P. Carmelo, P. D. Sacramento Phys. Rev. B 68, 085104 (2003)

\bibitem{carmelo2}
	J. M. P. Carmelo, K. Penc, P.D. Sacramento, R. Claessen, J. Phys. IV France 114 (2004) 45

\bibitem{carmelo3}
	J. M. P. Carmelo, J. M. Rom\'{a}n, K. Penc, Nucl. Phys. B 683(3):387-422

\bibitem{carmelo4}
	N. M. R. Peres, P D Sacramento, J M P Carmelo, J. Phys.: Condens. Matter 13(22):5135

\bibitem{carmelo5}
	J. M. P. Carmelo, J. M. Rom\'{a}n, K. Penc, Nucl. Phys. B 683 387 (2004)

\bibitem{carmelo6}
	J. M. P. Carmelo, L. M. Martelo, P. D. Sacramento, J. Phys.: Condens. Matter 16 (2004) 1375–1399

\bibitem{carmelo61}
	J. M. P. Carmelo, T. \v{C}ade\v{z}, Nuclear Physics B 904 (2016) 39–85

\bibitem{carmelo62}
	J. M. P. Carmelo, T. \v{C}ade\v{z}, Nuclear Physics B 914 (2017) 461–552

\bibitem{carmelo63}
	J. M. P. Carmelo, S. \"{O}stlund, M. J. Sampaio, Ann. Phys. 325 (2010) 1550.

\bibitem{carmelo64}
	J. M. P. Carmelo, P. D. Sacramento, Phys. Rep. Vol. 749, 27 July 2018, 1-90

\bibitem{carmelo7}
	M. Sing, U. Schwingenschl\"{o}gl, R. Claessen, P. Blaha, J. M. P. Carmelo, L. M. Martelo, P. D. Sacramento, M. Dressel, C. S. Jacobsen, Phys. Rev. B 68, 125111 (2003)

\bibitem{carmelo8}
	J. M. P. Carmelo et al 2006 J. Phys.: Condens. Matter 18 5191

\bibitem{carmelo9}
	J. M. P. Carmelo, K. Penc, L. M. Martelo, P. D. Sacramento, J. M. B. Lopes dos Santos, R. Claessen, M. Sing, U. Schwingenschl\"{o}gl, EPL 67 233 (2004)

\bibitem{carmelo10}
	J.M.P. Carmelo, F. Guinea, K. Penc, P.D. Sacramento, Physica B: Cond. Mat. Vol. 359–361 (2005), 1427-1429

\bibitem{rotel1}
	A.B. Harris, R.V. Lange, Phys. Rev. 157 295 (1967)

\bibitem{rotel2}
	A.H. MacDonald, S.M. Girvin, D. Yoshioka, Phys. Rev. B 37 9753 (1988)

\bibitem{Uinfty1}
	F. Woynarovich, J. Phys. C 15 85 (1982)

\bibitem{Uinfty2}
	F. Coll, Phys. Rev. B 9 2150 (1974)

\bibitem{penc1}
	K. Penc, F. Mila, H. Shiba, Phys. Rev. Lett. 75 894 (1995)

\bibitem{penc2}
	K. Penc, K. Hallberg, F. Mila, H. Shiba, Phys. Rev. Lett. 77 1390 (1996)

\bibitem{penc3}
	K. Penc, K. Hallberg, F. Mila, H. Shiba, Phys. Rev. B 55 15475 (1997)

\bibitem{penc4}
	J. Favand, S. Haas, K. Penc, F. Mila, E. Dagotto, Phys. Rev. B 55, R4859(R) (1997)

\bibitem{halff1}
	P. W. Anderson, Solid State Phys. 14 99 (1963)

\bibitem{others1}
	F. H. L. E{\ss}ler, V. E. Korepin, K. Schoutens, Phys. Rev. Lett. 67, 3848

\bibitem{others11}
	F. H. L. E{\ss}ler, V.E, Korepin, K. Schoutens, Nucl. Phys. B 372 559 (1992)

\bibitem{others12}
	F. H. L. E{\ss}ler, V.E, Korepin, K. Schoutens, Nucl. Phys. B 384 431 (1992)

\bibitem{others2}
	F. H. L. E{\ss}ler, V. E. Korepin, Nucl. Phys. B 426 505 (1994)

\bibitem{others3}
	T. Deguchi et al., Phys. Rep. 331 (2000) 197-281

\bibitem{carmelo94percent}
	J. M. P. Carmelo, K. Penc 2006 J. Phys.: Condens. Matter 18 2881

\bibitem{carmeloS}
	J. M. P. Carmelo, J. Phys.: Condens. Matter 17 5517 (2005)

\bibitem{josenuno2}
	J. M. P. Carmelo, N. M. R. Peres Phys. Rev. B 51, 7481

\bibitem{TTFTCNQ}
	H. Benthien, F. Gebhard, E. Jeckelmann, Phys. Rev. Lett. 92 256401 (2004)

\bibitem{pdt}
	J. M. P. Carmelo, K. Penc, D.Bozi, Nucl. Phys. B 725 421 (2005)

\bibitem{pdt2}
	J. M. P. Carmelo, https://arxiv.org/pdf/cond-mat/0405411.pdf

\bibitem{orthcat1} 
	P.W. Anderson, Phys. Rev. Lett 18 1049 (1967)

\bibitem{orthcat2}
	I. Affleck, Nucl. Phys. B 58 35 (1997)

\bibitem{JW}
	Y. R. Wang, Phys. Rev. B 46 (1992) 151

\end{thebibliography}
\end{document}